\documentclass[twocolumn,
	aps, prd,
	10pt, notitlepage, 
        floats, floatfix,
	amsmath, amssymb, amsfonts, eqsecnum,
	superscriptaddress,
	showpacs, showkeys,
	nofootinbib,
 	longbibliography,
]{revtex4-2}
\pdfoutput=1
\usepackage{amssymb}
\usepackage{amsmath}
\usepackage{epsfig}
\usepackage{hyperref}
\usepackage{breakurl}
\usepackage{multirow}
\usepackage{array}
\usepackage{bm}
\usepackage{color}
\usepackage{tikz}
\usepackage[utf8]{inputenc}
\usepackage{braket}
\usepackage{graphicx}
\usepackage{footmisc}

\usepackage{colortbl}
\definecolor{blue2}{cmyk}{1, 0.1, 0.1, 0}

\definecolor{pyBlue}{RGB}{31, 119, 180}
\definecolor{pyRed}{RGB}{214, 39, 40}
\definecolor{pyGreen}{RGB}{44, 160, 44}
\definecolor{pyBlue2}{RGB}{0, 111, 237}
\definecolor{pyRed2}{RGB}{224, 52, 36}

\usepackage{colortbl}
\definecolor{summersky}{cmyk}{0.71,0.33,0,0.5}
\definecolor{flamingo}{cmyk}{0,0.51,0.71,0.5}
\definecolor{rp}{cmyk}{0.2, 1, 0.6, 0}
\definecolor{pacificblue}{cmyk}{0.95,0.3,0, 0.5}
\definecolor{gray60}{cmyk}{0.4,0.4,0,0.8}

\newcommand{\red}[1]{\textcolor{pyRed}{#1}}
\newcommand{\blue}[1]{\textcolor{pyBlue}{#1}}

\newcommand{\Rlm}[4]{{}_{#1}R_{#2 #3 #4}}
\newcommand{\Rin}[4]{{}_{#1}R^{\rm in}_{#2 #3 #4}}
\newcommand{\sSlm}[3]{{}_{#1} S_{#2 #3}}

\newcommand{\rs}{r_*}
\newcommand{\Binc}[3]{B^{\rm inc}_{#1 #2 #3}}
\newcommand{\Bref}[3]{B^{\rm ref}_{#1 #2 #3}}

\renewcommand{\(}{\left(}
\renewcommand{\)}{\right)}
\renewcommand{\[}{\left[}
\renewcommand{\]}{\right]}

\usetikzlibrary{decorations.markings}
\usepackage{cleveref}
\crefformat{section}{\S#2#1#3}
\crefformat{subsection}{\S#2#1#3}
\crefformat{subsubsection}{\S#2#1#3}

\makeatletter
\def\simgt{\mathrel{\lower2.5pt\vbox{\lineskip=0pt\baselineskip=0pt
           \hbox{$>$}\hbox{$\sim$}}}}
\def\simlt{\mathrel{\lower2.5pt\vbox{\lineskip=0pt\baselineskip=0pt
           \hbox{$<$}\hbox{$\sim$}}}}

\def\sectionskip{\vskip .2 cm}

\makeatother

\def\spa#1.#2{\left\langle#1\,#2\right\rangle}
\def\spb#1.#2{\left[#1\,#2\right]}
\def\sand#1.#2.#3{%
\left\langle#1{\vphantom1}\right|{#2}\left|#3\right]}%
\def\sandmp#1.#2.#3{%
\left\langle#1{\vphantom1}\right|{#2}\left|#3\right]}%
\def\sandpm#1.#2.#3{%
\left[#1{\vphantom1}\right|{#2}\left|#3\right\rangle}%
\def\sandmm#1.#2.#3{%
\left\langle#1{\vphantom1}\right|{#2}\left|#3\right\rangle}%
\def\sandpp#1.#2.#3{%
\left[#1{\vphantom1}\right|{#2}\left|#3\right]}%

\renewcommand{\imath}{\mathrm{i}}

\def\Section#1{\noindent {\it #1}}
\newcommand{\be}{\begin{equation}}
\newcommand{\ee}{\end{equation}}

\def\S{{\mathbb S}}

\begin{document}

\title{Black Hole Perturbation Theory Meets CFT$_2$: \\[4pt]\fontsize{10}{0} \selectfont  Kerr  Compton Amplitudes from  Nekrasov-Shatashvili Functions}

\author{Yilber Fabian Bautista}
\email{yilber-fabian.bautista-chivata@ipht.fr}
\affiliation{Institut de Physique Théorique, CEA, Université Paris–Saclay,
F–91191 Gif-sur-Yvette cedex, France}
\author{Giulio Bonelli}
\email{bonelli@sissa.it}
\affiliation{International School of Advanced Studies (SISSA), via Bonomea 265, 34136 Trieste, Italy}
\affiliation{Institute for Geometry and Physics, IGAP, via Beirut 2, 34151 Trieste, Italy}
\affiliation{INFN Sezione di Trieste, via Valerio 2, 34127 Trieste, Italy}
\author{Cristoforo Iossa}
\email{Cristoforo.Iossa@unige.ch}
\affiliation{Section de Math\'{e}matiques, Universit\'{e} de Gen\`{e}ve, 1211 Gen\`{e}ve 4, Switzerland}
\author{Alessandro Tanzini}
\email{tanzini@sissa.it}
\affiliation{International School of Advanced Studies (SISSA), via Bonomea 265, 34136 Trieste, Italy}
\affiliation{Institute for Geometry and Physics, IGAP, via Beirut 2, 34151 Trieste, Italy}
\affiliation{INFN Sezione di Trieste, via Valerio 2, 34127 Trieste, Italy}
\author{Zihan Zhou}
\email{zihanz@princeton.edu}
\affiliation{Department of Physics, Princeton University, Princeton, NJ 08540, USA}

\begin{abstract}
We present a novel study of Kerr Compton amplitudes in a partial wave basis in terms of  the  Nekrasov-Shatashvili (NS) function of the \textit{confluent Heun equation} (CHE). 
Remarkably,  NS-functions enjoy analytic properties and symmetries that are naturally inherited by the Compton amplitudes. Based on this, we characterize the  analytic dependence of the Compton phase-shift in the Kerr spin parameter and provide a direct comparison to the standard post-Minkowskian (PM) perturbative approach within General Relativity (GR). We also analyze the universal large frequency behavior of the relevant
characteristic exponent of the CHE -- also known as the renormalized angular momentum -- and find agreement with numerical computations.
Moreover, we  discuss the analytic continuation in the harmonics quantum number $\ell$ of the partial wave, and show that the limit to the  physical integer values 
commutes with the PM 
expansion of the observables.
Finally, we obtain the contributions to the tree level,  point-particle,  gravitational   Compton amplitude in a covariant basis through $\mathcal{O}(a_{\text{BH}}^8)$,
without the need to  take the super-extremal limit for  Kerr spin. 
\end{abstract}

\maketitle

\section{Introduction}

The study of black hole perturbation theory  has seen a resurgence in recent years after the observation of the gravitational waves generated by the coalescence of binary black holes \cite{LIGOScientific:2016aoc,KAGRA:2020agh}.
This revitalization has led to the development of novel perturbative approaches for examining black holes' responses to external perturbations. These methods draw heavily from Quantum Field Theory (QFT)-inspired techniques, including (quantum) worldline effective field theory (EFT) \cite{Goldberger:2004jt,Rothstein:2014sra,Goldberger:2022rqf,Goldberger:2022ebt,Porto:2016pyg,Kalin:2020mvi,Mogull:2020sak},
on-shell amplitudes 
\cite{Cheung:2018wkq,Kosower:2018adc,Bern:2019crd,Bern:2020buy,Buonanno:2022pgc,Cheung:2023lnj,Kosmopoulos:2023bwc,Bjerrum-Bohr:2022blt,Brandhuber:2023hhy,Herderschee:2023fxh,Elkhidir:2023dco,Georgoudis:2023lgf,Caron-Huot:2023vxl}, and the effective-one-body (EOB) approximation \cite{Buonanno:1998gg,Buonanno:2000ef}. A crucial aspect of these approaches is to match the physical observables derived from effective models with those calculated in General Relativity (GR), which is key to identifying unknown parameters within the effective theories. Therefore, it is important to exactly solve the differential equations in GR, 
as well as providing organizing principles to interpret the mathematical results.

This work aims to establish a connection between a novel computational approach to solve the Teukolsky master equation (TME) and the analysis of Compton scattering amplitudes in a Kerr black hole (KBH) background. This computational scheme is grounded in transforming the separated radial and angular components of the TME into a second-order ordinary differential equation (ODE), notably the \textit{confluent} Heun equation (CHE) \cite{Ronveaux1995-ra}. This transformation allows to relate the solutions of the equation to {\it classical} Virasoro conformal blocks, as detailed in \cite{Bonelli:2021uvf}.
By exploiting the known analytic properties of these conformal blocks and their representation through the Nekrasov-Shatashvili (NS) special function \cite{Nekrasov:2009rc}, new explicit solutions for the connection coefficients of the CHE could be derived \cite{Bonelli:2022ten}. 
This method has already been applied to the study of physical observables in a variety of gravitational backgrounds including (A)dS black holes \cite{Dodelson:2022yvn,Aminov:2023jve}, fuzzballs \cite{Bianchi:2021xpr,Bianchi:2022qph,Bianchi:2023sfs,Giusto:2023awo}, and the  astrophysically relevant KBH, with the first application appearing in the context of the exact computation of the spectrum of Quasi Normal Modes \cite{Aminov:2020yma}, and more recent approaches to compute greybody factor,  Love numbers \cite{Bonelli:2021uvf,Consoli:2022eey} and the study of the post-Newtonian (PN) dynamics in the two-body problem \cite{Fucito:2023afe}.

In this letter, we show  that Kerr Compton amplitudes written  in a partial wave basis can be directly expressed in terms of the NS-function.
As a consequence, the analytic properties of the NS-function translate into sharp statements for the Compton scattering phase-shift. This allows us to:
\begin{itemize}
\item Non-perturbatively characterize  the polynomial dependence on the KBH spin parameter of different contributions to the phase-shift directly related to the NS-function.
\item By comparing to the more traditional method of 
Mano, Suzuki, and Takasugi (MST)
\cite{Mano:1996mf,Mano:1996vt,Mano:1996gn,Sasaki:2003xr}, to resum the perturbative -- post-Minkowskian (PM)  -- expansion into exponential functions of derivatives of the NS-function.
\item To  study analytically the large frequency behaviour of the MST renormalized angular momentum  and compare the results  to  numerical predictions. 
\end{itemize}

By studying the Compton phase-shift 
in a PM-fashion -- namely $\epsilon \equiv 2 G M \omega \ll 1$ -- we find it naturally separates into a dominating and a depleted contribution.
This hierarchical distinction aligns  with the ``far zone" -- conservative, point-particle, leading contribution --
and ``near-zone" -- horizon completion -- factorization recently proposed in \cite{Ivanov:2022qqt,Saketh:2023bul}. 
The near-far factorization is well defined for harmonics of  generic-$\ell$ values, i.e. when analytically continue from $\ell \in \mathbb{N}$ to $\ell \in \mathbb{C}$. This continuation gives rise to apparent divergences once the physical limit $\ell \in \mathbb{N}$ is taken. In an MST language, this manifests as integer $\ell$-poles in the MST coefficients \cite{Casals:2015nja,Casals:2016soq}. In this work we show such poles are spurious and get canceled when adding the PM-expanded near and far zone contributions of the phase-shift together. 
The final results in this \textit{generic$-\ell$ prescription} agree  with  the ones computed in a \textit{fixed-$\ell$ prescription}, i.e. by solving the ODEs  staring with $\ell \in \mathbb{N}$ before PM-expanding \cite{Casals:2015nja,Casals:2016soq}. Therefore, we conclude the PM- and the  $\ell$-expansions actually commute.
Finally, we provide  a new interpretations of the results presented in \cite{Bautista:2022wjf} for the  higher-spin, tree-level Gravitational Compton amplitude in terms of only far-zone physics, while expanding the state of the art results to eight-order in the Kerr-spin multipole expansion. As mentioned,  this far zone computation corresponds to the  point-particle limit of the BH, while being purely conservative and polynomial in the KBH spin parameter $\chi$; therefore, no analytic continuation in $\chi$ is required.

\section{ spin-$s$ perturbations off Kerr }

The radiative content for perturbation of  spin-weight $s$  off a  Kerr black hole (KBH) of mass $M$ and spin $a_{\text{BH}}$ is fully encoded in the Teukolsky  scalar  ${}_{s}\psi$, which solves TME. As shown by  Teukolsky's seminal work \cite{teukolsky1972rotating,teukolsky1973perturbations,teukolsky1974perturbations},  ${}_{s}\psi$  admits separation of variables in the frequency domain. Using   $(t,r,\vartheta,\varphi)$ as the Boyer-Lindquist coordinates it can be  explicitly  expressed as 
\begin{align}
\label{eq:TME}
{}_{s}\psi(t,r,\vartheta,\varphi)=\sum_{\ell m}\int d\omega \, e^{-i \omega t}\Rlm{s}{\ell}{m}{}(r)\sSlm{s}{\ell}{m}(\vartheta,\varphi; a_{\text{BH}} \omega)\,.
\end{align}
Here $\Rlm{s}{\ell}{m}{}(r)$ solves the radial Teukolksy equation (RTE), whereas $\sSlm{s}{\ell}{m}(\vartheta,\varphi;a_{\text{BH}} \omega)$ correspond to the spin-weighted spheroidal harmonics. As mentioned above, both the RTE and the angular equation can be reduced to CHE  after a suitable change of variables. The RTE has singularities at the inner and outer horizons of the KBH, and at the  boundary at infinity. More broadly, Teukolsky equations for a generic class of Type-D space-times  correspond to Heun's equations  of certain type, classified by the structure of their singular point \cite{Batic:2007it}.

In this work we consider plane wave   perturbations off KBH imposing the physical boundary conditions for the radial function to be purely ingoing at the BH horizon and a superposition of an incoming and a reflected wave at future null infinity (see \cref{fig:matching}),
\begin{equation}\label{eq:bdary}
\begin{split}
	\Rin{s}{\ell}{m}{}(r)&= \Delta^{-s}e^{-i\tilde{\omega} \rs}\,,\qquad\qquad\qquad\quad\rs\to-\infty\,, \\
 \Rin{s}{\ell}{m}{}(r)&= {}_s\Binc{\ell}{m}{} \frac{ e^{-i\omega \rs}}{r}{+}{}_s\Bref{\ell}{m}{} \frac{e^{i\omega \rs}}{ r^{(2s+1)}} \,,\quad \rs\to \infty\,.
\end{split}
\end{equation}
Here $\tilde \omega=\omega-\frac{m \chi}{2 r_+}$  is the co-rotating  frequency, $\chi = a_{\text{BH}}/(GM)$ is the dimensionless spin of the KBH, $r_\pm=GM(1\pm\kappa)$ are the roots of $\Delta=r^2+ 2GM r +a_{\rm BH}^2$, and $\kappa=\sqrt{1-\chi^2}$ . The tortoise coordinate $\rs$ is determined from the differential equation $\frac{d\rs}{dr}=\frac{r^2+a_{\text{BH}}^2}{\Delta}$  \cite{Sasaki:2003xr}. 

The main objects of interest are the (Compton) scattering phase-shift ${}_s\delta_{\ell m}^P$ and the absorption probability ${}_s\eta_{\ell m}^P$,   which are fully  determined from the asymptotic behaviour of the radial functions  \cite{futterman88}:
\begin{equation}\label{eq:phase-shift-gen}
{}_s\eta_{\ell m}^P e^{2 i {}_s \delta_{\ell m}^P} = (-1)^{\ell + 1} \frac{{}_s B_{\ell m}^{\text{ref}}}{{}_s B_{\ell m}^{\text{inc}}} \times (2 \omega)^{2s} A_s^{P} ~,
\end{equation}
being $A_s^P$  a function of the Teukolsky-Starobinsky constant (see \eqref{eq:teukstarovins}), with $P$ a parity label. ${}_s\Binc{\ell}{m}{}$ and ${}_s\Bref{\ell}{m}{}$ are  called connection coefficients of the CHE since they allow us to express a local solution close to a singular point in terms of a local basis of solutions centered around a different singular point. The Heun connection coefficients for generic boundary conditions have been explicitly computed in \cite{Bonelli:2022ten} (see appendix B for a review of the derivation). Using these results we find

\begin{widetext}
\begin{equation}\label{eq:bbnek}
\begin{aligned}
 \frac{{}_s B_{\ell m}^{\text{ref}}}{{}_s B_{\ell m}^{\text{inc}}} = - i \frac{e^{2i \epsilon (\log(|2\epsilon|)-1/2)}}{|2\omega|^{2s}} e^{\partial_{m_3} F - \frac{L}{2} } \frac{\sum_{ \sigma = \pm} \frac{\Gamma\left(1{-}2\sigma a \right) \Gamma\left({-}2\sigma a \right)({-}L)^{\sigma a} e^{{-}\frac{\sigma}{2} \partial_a F}}{ \Gamma\left(\frac{1}{2}{-}\sigma a{+}m_1\right)\Gamma\left(\frac{1}{2}{-}\sigma a{+}m_2\right)\Gamma\left(\frac{1}{2}{-}\sigma a{+}m_3\right)} }{ \sum_{ \sigma'{ =} \pm} \frac{ \Gamma\left(1{-}2\sigma' a\right) \Gamma\left({-}2\sigma' a\right)L^{\sigma' a} e^{{-}\frac{\sigma'}{2} \partial_a F}}{ \Gamma\left(\frac{1}{2}{-}\sigma' a{+} m_1 \right)\Gamma\left(\frac{1}{2}{-}\sigma' a {+} m_2 \right)\Gamma\left(\frac{1}{2}{-}\sigma' a-m_3\right)}}\,,
\end{aligned}
\end{equation}
\end{widetext} 
with dictionary of parameters \cite{Aminov:2020yma}
\begin{equation}\label{eq:dict}
\begin{aligned}
&m_1 {=} i \frac{m\chi-\epsilon}{\kappa} \,, \quad m_2 {=} -s {-} i \epsilon \, , \quad m_3 {=}  i \epsilon{-}s \,,\quad
 L {=} -2i \epsilon\kappa \,, \\
&u = -\lambda - s(s+1) + \epsilon (  i s\kappa- m \chi) 
 + \epsilon^2 (2 + \kappa ) \,,
\end{aligned}
\end{equation}
and  $\lambda$  the  spheroidal eigenvalue. $a$ is implicitly determined from the so called Matone relation \cite{Flume:2004rp,Matone:1995rx}:
\begin{equation}\label{eq:Matone}
u = \frac{1}{4}-a^2+L \partial_L F \left(m_1,m_2,m_3,a,L\right) \,.
\end{equation}
All the complexity in computing \eqref{eq:bbnek} is then hidden  in the special function $F(m_1,m_2,m_3,a,L)$. This is a so called NS-function, and it is given as a convergent series\footnote{For the convergence of Nekrasov partition functions see \cite{Arnaudo:2022ivo}.} in $L$ whose coefficients are given explicitly in terms of combinatorial formulas (see appendix B for concrete formulas). NS-functions are a class of special functions which appeared for the first time in the context of $\mathcal{N}=2$ supersymmetric gauge theories and Liouville CFT\footnote{In the gauge theory context, $F$ appears as the instanton partition function of a $\mathcal{N}=2$ SU(2) gauge theory with 3 hypermultiplets of masses $m_1,m_2,m_3$. $L$ is the instanton counting parameter and $a$ the Cartan vacuum expectation value in the Coulomb branch. $a$ can also be understood as the quantum A-period of the CHE. In the Liouville CFT, $\sigma=\pm$ denotes two different intermediate dimensions in the conformal block expansion.} \cite{Nekrasov:2009rc,Alday:2009aq}. Different NS-functions make their appearance in the connection problem of  Heun equations of different types \cite{Bonelli:2022ten}. Since in this paper we are only dealing with the CHE, we will not make a notational effort to distinguish the NS-function $F$ from its siblings.
From \eqref{eq:dict}, it follows that
$L$ acts effectively as a PM-parameter, aligning the $L$ expansion of $F$ with the standard PM-expansion used by the MST method. 
\footnote{An similar observation was made recently in  \cite{Fucito:2023afe} in  the post-Newtonian context.}
This observation is crucial and  allows  for a direct comparison of the two methods as we will see below.

For practical purposes, our strategy to compute \eqref{eq:bbnek} is the following: 
\begin{itemize}
\item compute $F$ up to order $L^{n_{\text{max}}}$,
\item invert \eqref{eq:Matone} perturbatively in $L$ to obtain $a(m_i,u,L)$,
\item plug $a(m_i,u,L)$ back in $F$ and evaluate \eqref{eq:bbnek} substituting the dictionary \eqref{eq:dict} up to order $\epsilon^{n_{\text{max}}}$.
\end{itemize}
We include the explicit expression for $F$ up to $\mathcal{O}(L^9)$ in the  ancillary files  of this work \cite{ancill}. For concreteness, at leading order one finds 
\begin{equation}\label{eq:concretea}
a = -\frac{\sqrt{(1+2s)^2+4\lambda}}{2}+\mathcal{O}(\epsilon) = -\frac{1}{2}-\ell+\mathcal{O}(\epsilon)^2 \,.
\end{equation}

Note that in the $\epsilon \ll 1$ limit $|L|^{a} \sim e^{-(1+2\ell)\log(\epsilon)/2}$, therefore 
only $\sigma= 1$ terms contributes to the sums  in \eqref{eq:bbnek} at leading order. We call this the {\it far zone} contribution. The $\sigma = -1$ terms are thus suppressed by a factor of $|L|^{-2a} \sim \epsilon^{1+2\ell}$, which coincides with the order at which BH horizon effects start to become relevant \cite{Page:1976df,Page:1976ki,Saketh:2022xjb,Saketh:2023bul}. For this reason, we call the factor containing these terms the {\it near zone}. 
We therefore rewrite formula \eqref{eq:bbnek} in the more familiar form 
\begin{widetext}
\begin{equation}\label{eq:nearpfarCFT}
    \frac{{}_s B_{\ell m}^{\text{ref}}}{{}_s B_{\ell m}^{\text{inc}}} = {\red{\underbrace{ \frac{e^{2i \epsilon( \log(|2\epsilon|)-1/2)}}{|2\omega|^{2s}} e^{\partial_{m_3} F - \frac{L}{2} } e^{ i\pi \left(a - \frac{1}{2}\right)  }\frac{\Gamma\left(\frac{1}{2}- a-m_3\right)}{\Gamma\left(\frac{1}{2}- a+m_3\right)}}_{\rm far \, zone}}} \times {\blue{\underbrace{\frac{1 + e^{-i \pi a} \mathcal{K}}{1 + e^{i\pi a} \frac{\cos(\pi(m_3 -a))}{\cos(\pi(m_3+a))} \mathcal{K}}}_{\rm near \, zone}}} ~,
\end{equation}
where
\begin{equation}\label{eq:Kappa}
    \mathcal{K} = |L|^{-2 a} \frac{ \Gamma(2 a) \Gamma(2 a +1) \Gamma\left(m_3-a +\frac{1}{2}\right) \Gamma\left(m_2-a +\frac{1}{2}\right) \Gamma\left(m_1-a +\frac{1}{2}\right)}{ \Gamma(-2 a) \Gamma(1-2 a) \Gamma\left(m_3+a +\frac{1}{2}\right) \Gamma\left(m_2+a +\frac{1}{2}\right) \Gamma\left(m_1+a +\frac{1}{2}\right)} e^{\partial_a F}\,,
\end{equation}
\end{widetext}
thus resembling the near-far factorization proposed in \cite{Ivanov:2022qqt} for the coefficient-ratio written in the MST-language (see \eqref{eq:factorization} below). The function $\mathcal{K}$ here is called the tidal response function and we shall give a more detailed explanation on it in Appendix A. From the above equation, we observe that the far zone contribution can be written as an analytic functions of $L$ while the near zone is non-analytic in $L$ for generic values of $a$,  hence sharing an analog of the analytic structure in $G$ discussed in \cite{Saketh:2023bul}. 
A detailed comparison with MST method will be done in the next section.

\section{NS-Function, MST  PM-Resummation,  and the High Frequency limit}

Since the phase-shift \eqref{eq:phase-shift-gen} depends on the  NS-function via the connection formula \eqref{eq:nearpfarCFT}, the symmetry properties of $F$ are naturally imprinted in 
the Compton scattering amplitudes. We start this section then  by presenting some properties of the NS-function (see appendix B for conventions and derivations).

{\it Properties of the NS-function}--- 
The function 
\begin{equation}\label{eq:f-tilde}
\tilde{F}(m_1,m_2,m_3,a,L) = F(m_1,m_2,m_3,a,L)-\frac{m_3 L}{2}
\end{equation}
is invariant under permutations of $(m_1, m_2, m_3)$ and under the reflection $(m_i, L) \to (-m_i, -L)$. Accordingly, it only depends on combinations that are left invariant under such transformations, that is 
\begin{equation}\label{eq:invcomb}
\begin{aligned}
&(m_1 m_2 m_3) L \,, \,\, (m_1 m_2 m_3)^2 \,, \,\, L^2 \,, \,\, \sum_{i=1}^3 m_i^{2n} \,, \\ 
&\,\quad (m_1 m_2)^{2n}+(m_2 m_3)^{2n}+(m_1 m_3)^{2n} \,,
\end{aligned}
\end{equation}
with $n \in \mathbb{N}$. Furthermore, if $\tilde{F}$ is Taylor expanded in $L$, 
\begin{equation}
\tilde{F}=\sum_n c_n(m_1,m_2,m_3,a) L^n \,, 
\end{equation}
then
\begin{equation}\label{eq:coeff}
\lim_{m_i\to\infty} c_n(m_1,m_2,m_3, a) \propto m_i^n \,.
\end{equation}
Substituting the dictionary \eqref{eq:dict} in \eqref{eq:invcomb}, we note that $\tilde{F}$ can only depend on $\kappa^2 = 1-\chi^2$, so, by \eqref{eq:coeff}, it follows that factors of $\kappa^2$ can only appear in the numerator of $c_n$'s. This therefore proves that $\tilde{F}$ depends only  polynomially on the  spin $\chi$. Moreover, since all the invariant combinations in \eqref{eq:invcomb} are real after substituting the dictionary \eqref{eq:dict}, we see that $\mathcal{\tilde{F}}$ is real at all orders in $\epsilon$. Subtracting $m_3 L /2$ in the LHS and RHS of \eqref{eq:Matone} we see that $a^2$ is also real \footnote{In the small frequency expansion, $a$ is real.} and depends polynomially on $\chi$.
As a consequence, we conclude that the
{\it  far-zone phase-shift is polynomial in terms of BH spin.} This polynomial structure aligns with spin-induced multipole expansion used in the worldline EFT \cite{Levi:2015msa}.

Let us also comment on the dependence of the NS-function on $a$. $F$ is invariant under $a \to -a$, as indicated in \eqref{eq:bbnek} and \eqref{eq:Matone}. Moreover, $F$ has poles at $a = \pm n/2$ for $n\in \mathbb{N}$ \cite{Gorsky:2017ndg}. Simple poles at $a=\pm n/2$ appear at order $L^n$, and poles of higher orders appear at higher orders in the $L$-expansion.

{\it NS-function and PM-Resummation}--- In the  PM approach, it is  customary   to use the MST method for solving the TME \cite{Mano:1996gn,Mano:1996mf,Mano:1996vt,Sasaki:2003xr}. In this approach one matches the asymptotic solutions converging in the near ($r_+\le |r| <\infty$) and far ($r_+<|r| \le \infty$) zone perturbatively in $\epsilon$ after imposing the boundary condition \eqref{eq:bdary} (see \cref{fig:matching}). In doing so, one introduces the so called renormalized angular momentum $\nu(s,\ell,m,\omega)$, and the MST coefficients $a_n^\nu$, which are computed perturbatively in $\epsilon$ from a three-term recursion relation that is required by the convergence condition. The connection coefficients ${}_s\Binc{\ell}{m}{}$ and ${}_s\Bref{\ell}{m}{}$ are then expressed in terms of infinite sums involving $a_n^\nu$ and $\nu$ \cite{Sasaki:2003xr}. We refer the reader to Appendix A for a review of the MST method\footnote{ 
For recent mathematical results on the perturbative expansion of the connection formulae for CHE see instead
\cite{Lisovyy:2022flm}.}.

Comparing to the MST solutions, the CFT results suggest the resummation of the MST sums for the  far-zone and near-zone respectively  \footnote{Indeed, this follows since  $a_{n}^\nu \sim \epsilon^{|n|} \sim (G M \omega)^{|n|} $ in generic-$\ell$ prescription as we will see below. Then, perturbative calculation using $a_n^\nu$ is equivalent to the  PM-expansion of $F$. \label{fn:PMMST}}
\begin{equation}\label{eq:rational}
\boxed{
e^{\partial_{m_3}F} {=} \frac{
\sum_{n=-\infty}^{+\infty}(-1)^n{(\nu+1+s-i\epsilon)_n\over 
(\nu+1-s+i\epsilon)_n}a_n^\nu}{\sum_{n=-\infty}^{+\infty}a_n^\nu},\,\,\, e^{\partial_a F} {=}\frac{ X_{-\nu-1} }{ X_\nu }\,,}
\end{equation}
where the non-trivial $X_\nu$ sums are given in \eqref{eq:X-MST}, whereas  the renormalized angular momentum is found to be 
\begin{equation}\label{eq:alphavsnu}
\boxed{
a = -\frac{1}{2}-\nu}\,.
\end{equation}
We have checked formulas \eqref{eq:rational} and \eqref{eq:alphavsnu}  hold up 9-PM order, and we expect them to hold true to all orders in perturbation theory, for generic spin-weight $s$, angular  momentum $\ell$, and azimuthal number $m$.  Indeed, we expect it would be possible to analytically prove this formula possibly along the lines of \cite{Lisovyy:2022flm}.

\begin{figure}[t!]
    \centering
    \includegraphics[scale = 0.4]{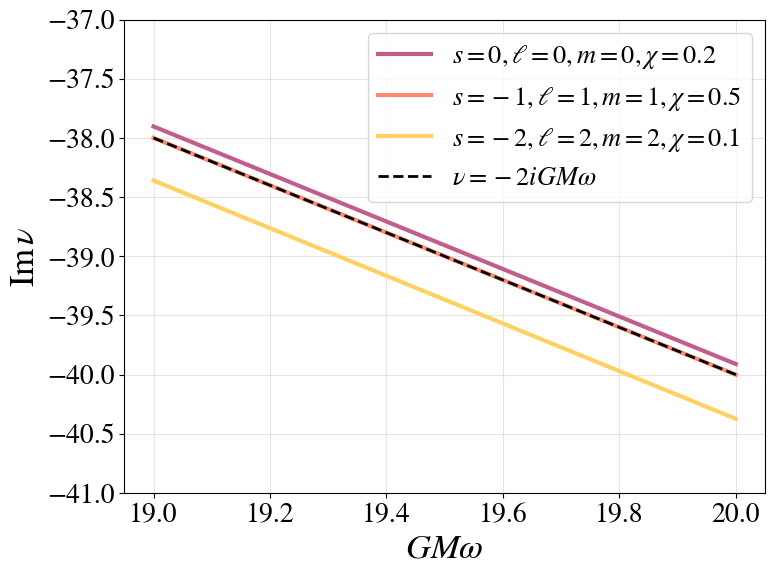}
    \caption{Numerical evaluation of the high frequency behavior of renormalized angular momentum $\nu$ using Black Hole Perturbation Toolkit \cite{BHPToolkit}. The solid lines represent $\nu$ for various perturbation parameters. The dashed line is the analytic estimation from \eqref{eq: analytic estimation nu}.}
    \label{fig:high_freq_nu}
\end{figure}

{\it Towards High Frequency Scattering}--- In some corners of the parameter space, the perturbative series in $L$ -- which at the same time  defines $F$ --  simplifies drastically. An interesting example  where  such a  simplification takes place   is  the high frequency limit; that is,  when $G M \omega \gg 1$, which corresponds to 
\begin{equation}\label{eq:specialcase}
m_3+m_2\simeq 0 \,, \quad u \simeq \frac{1}{4}-m_3^2+\frac{L}{2}(m_3-m_1) \,.
\end{equation} 
As shown in detail in Appendix C, in this limit the NS-function reduces to
\begin{equation}\label{eq:Fhfl}
\tilde{F} \simeq - \frac{L m_1}{2} \,.
\end{equation}
A direct consequence from the Matone relation 
 \eqref{eq:Matone}, in conjunction with \eqref{eq:specialcase} and \eqref{eq:Fhfl}, 
 is therefore that  $a \simeq m_3$, hence one finds using \eqref{eq:alphavsnu}
that
\begin{equation}
\label{eq: analytic estimation nu}
a \simeq 2 i G M \omega \, \Rightarrow \nu \simeq -2 i G M \omega \,\, \text{as} \,\, G M\omega\to\infty \,.
\end{equation}
In  \cref{fig:high_freq_nu} we test this expression against numerical predictions. The above relation we find is universal for all $s,\ell,m,\chi \ll G M \omega$, a pattern verified in \cref{fig:high_freq_nu}.

\section{Generic-$\ell$ vs fixed-$\ell$ prescriptions}

Physical angular momentum $\ell$ attain positive integers values with $\ell \geq |s|$. However, when we fix $\ell = n \in \mathbb{N}$, from \eqref{eq:concretea} we observe $a$ becomes a half-integer at leading order in $\epsilon$. This complicates the structure of the PM expansion, since we are expanding close to the poles of the NS-function and of the gamma functions in \eqref{eq:Kappa}. Considering for example the case $n=0$ ($s=\ell=m=0$), $a \simeq -1/2$,  the leading divergence in terms of $a$ in the NS-function can be written as \cite{Gorsky:2017ndg}
\begin{equation}\label{eq:leadindiv}
    F \sim \sum_{k=1}^\infty \frac{(m_1 m_2 m_3 L)^k}{(2a+1)^{2k-1}} = \mathcal{O}(\epsilon^2)\,.
\end{equation} 
Substituting the Kerr dictionary \eqref{eq:dict}, we see   no $1/\epsilon$ singularities actually appear in $F$
since the residue of the $2a=-1$ pole cancels such   divergences. However, it is crucial to notice that all terms with $k\geq 1$ in the $L$ series in \eqref{eq:leadindiv} will contribute to the $\epsilon^2$ order. In this sense,  the naive $L$-expansion no longer coincide with the PM expansion, and a resummation is needed.
A similar argument can be made for the divergent gamma factors  entering in 
\eqref{eq:Kappa}, and  for any $s,\ell,m$. Note that these complications appear  when one   hits the poles of the NS-function, that is, after $\mathcal{O}(\epsilon^{2\ell+1})$: no integer-$\ell$ issue   will arise before the horizon effects start to become important.

To avoid this kind of difficulties, we instead follow the route of analytically continue from $\ell \in \mathbb{N}$ to $\ell \in \mathbb{C}$, performing the low-frequency expansion in $L \sim \epsilon \ll 1$ and  going back to the physical limit  $ \ell \in \mathbb{N}$ only  at the final stage of the computations: we dub this approach the \textit{generic-$\ell$ prescription} \cite{Kol:2011vg}.

We  use the generic-$\ell$  prescription at the level of the phase-shift \eqref{eq:phase-shift-gen}. Let us then  comment on the structure of ${}_s\delta_{\ell m}^P$  in this approach. 
For starters, even though $\tilde{F}$ is purely real, $\partial_{m_3} \tilde{F}$ in \eqref{eq:nearpfarCFT} breaks the symmetries of the invariants combinations listed in \eqref{eq:invcomb}, and therefore it contains both real and imaginary parts. It is desirable to separate the real and imaginary contributions to ${}_s\delta_{\ell m}^P$ as they are associated to conservative and dissipative effects respectively.  
Interestingly, based on the conservation of energy flux between  infinity and the horizon, the identity  (see Eq.(68) in Ref.~\cite{Mano:1996gn})
\begin{equation}\label{eq:flux}
    e^{{\rm Re}[\partial_{m_3} F]} \Bigg| \frac{\Gamma\(\frac{1}{2} - a - m_3\)}{\Gamma\(\frac{1}{2} - a + m_3\)} \Bigg| = |A_s^P|^{-1} ~, s<0 ~, a \in \mathbb{R} ~.
\end{equation}
combined with Eq.~\eqref{eq:phase-shift-gen}, allows to  conclude that in the low-frequency limit, the far zone scattering is purely elastic whereas only the near zone contributes to the absorption probability $\Gamma \sim \sum_{P} [1- ({}_s\eta_{\ell m}^{P})^2$] \cite{Saketh:2022xjb,Saketh:2023bul}. 
Indeed, by combining \eqref{eq:flux} with  Eq.~\eqref{eq:nearpfarCFT} and Eq.~\eqref{eq:phase-shift-gen}, one can straightforwardly get the  far and near contributions to the phase-shift 
\begin{equation}\label{eq:phase-far}
\begin{split}
    {}_s \delta_{\ell m}^{P,\rm FZ} &=  \underbrace{ \frac{1}{2} {\rm Im} \[\partial_{m_3} F \] {-} \frac{1- \kappa}{2} \epsilon+\frac{1}{2}{\rm Arg}[A_s^P]}_{\rm rational}   {+} \underbrace{\epsilon \log(2 |\epsilon|)}_{\rm tail}\\ &
    +\underbrace{\frac{1}{2} {\rm Arg} \[ \frac{\Gamma(\frac{1}{2} {-} a {-} m_3)}{\Gamma(\frac{1}{2} {-} a {+} m_3)}\]{+} \frac{\pi}{2}\(\ell {+} \frac{1}{2} {+} a \)}_{\rm transcendental}~,
\end{split}
\end{equation}
\begin{equation}\label{eq:phase-near}
    {}_s \delta_{\ell m}^{P, \rm NZ} = \frac{1}{2} {\rm Arg} \left[ \frac{1 + e^{-i \pi a} \mathcal{K}}{1 + e^{i\pi\alpha} \frac{\cos(\pi(m_3 - a))}{\cos(\pi(m_3+a))} \mathcal{K}} \right] ~,
\end{equation} 
and
\begin{equation}
    {}_s\eta_{\ell m}^{P}=\Bigg| \frac{1 + e^{-i \pi a} \mathcal{K}}{1 + e^{i\pi\alpha} \frac{\cos(\pi(m_3 - a))}{\cos(\pi(m_3+a))} \mathcal{K}}  \Bigg| ~.
\end{equation}
In Eq.~\eqref{eq:phase-far}, we also observe that the NS-function $F$ only contributes to the rational function of $\ell$ while the transendental contributions come from the ratio of gamma function and the additional $\pi$ factor. The logarithmic tail terms represent the imprints from the long-range Newtonian potential.

The drawback of the generic $\ell$ prescription is that when taking   the physical limit $\ell \rightarrow \mathbb{N}$ at a given order in $\epsilon$, we encounter spurious poles at $\ell \in \mathbb{N}$ \cite{Casals:2015nja,Casals:2016soq} both in the near  and far zones \footnote{Note that, in the low-frequency expansion, $\epsilon \ll 1$, the poles $\ell \in \mathbb{N}$ come from the Taylor expansion of $F$ in Eq.~\eqref{eq:phase-far} and the ratio of Gamma functions in Eq.~\eqref{eq:phase-near}.}.
Consider for instance the $s=0$ perturbation at order $\epsilon^4$ ; its  exhibits the specific pole structure\footnote{Parity, $P=0$ for perturbations of spin-weight $s<2$, we therefore drop the $P$ label in this example.}:
\begin{equation}
\begin{aligned}
\label{eq: generic ell}
    {}_0 \delta_{\ell 1}^{\rm FZ}|_{\ell \rightarrow 1} & = \frac{\chi}{72(\ell-1)} \epsilon^4 + {\rm const} ~, \\
    \quad {}_0 \delta_{\ell 1}^{\rm NZ} |_{\ell \rightarrow 1} & = - \frac{\chi}{72(\ell -1)} \epsilon^4 + {\rm const} ~.
\end{aligned}
\end{equation}
To avoid these poles,  the traditional  MST approach uses a 
{\it fixed-$\ell$ prescription}, i.e. fixing $\ell,m \in \mathbb{N}$ before solving the Teukolsky equation \cite{Casals:2015nja,Casals:2016soq}.  Intriguingly, as Eq. \eqref{eq: generic ell} demonstrates, the poles in the near-zone precisely cancels those in the far zone, a pattern we have verified up to the $\mathcal{O}(\epsilon^8)$, for generic $s,\ell,m$. 
This cancellation suggests that the poles encountered when $\ell \rightarrow \mathbb{N}$ are essentially unphysical, and thus should cancel in any physical interpretation. A similar cancellation in an example of connection formula for hypergeometric function was pointed out in Ref.~\cite{Kol:2011vg}. For illustrative purposes,  in Appendix A we include  a explicit example keeping  the  non-divergent terms contributing in \eqref{eq: generic ell}. Moreover, a detailed comparison of the constant piece shows that when adding near-zone and far zone together, the results obtained using the  generic- and fixed-$\ell$ prescriptions completely agree with each other; we thus propose,
\begin{equation}
    \left({}_s \delta_{\ell m}^{P,\rm FZ} + {}_s \delta_{\ell m}^{P,\rm NZ} \right) \Big|_{\ell \rightarrow \mathbb{N}} = {}_s\delta_{(\ell \in \mathbb{N}) m}^{P,\rm FZ} + {}_s\delta_{(\ell \in \mathbb{N}) m}^{P,\rm NZ} ~,
\end{equation}
indicating that the two limit $\ell \rightarrow \mathbb{N}, \epsilon \rightarrow 0$ actually commute. In the remaining of this paper  we  only use the generic-$\ell$ formulation. We also stress here that the near-far factorization is only well-defined in the generic $\ell$ prescription. In the fixed-$\ell$ prescription, it is ill-defined and applying it leads to the effect of propagating non-Kerr particles in the $s-$ and $u-$ channels of the  Compton amplitude, as observed in Eq.~$(4.19)$ in \cite{Bautista:2022wjf}. This is because for $\ell \in \mathbb{N}$ when expanded in $\epsilon$, the two contributions   $\sigma=\pm 1$ in \eqref{eq:bbnek} mix with each other, i.e. $L^{a}$ and $L^{-a}$ both scale has half-integer scaling power, and hence one cannot separate the far-zone, $\sigma=-1$ piece from that.

The generic $\ell$ prescription also makes manifest of the locality structure of the scattering potential. This can be seen by fixing $\omega$ such that $\epsilon <1$ and taking $\ell \rightarrow \infty$, where the near zone and far zone phase-shift take following form respectively
\begin{equation}
    {}_s\delta_{\ell m}^{\rm FZ} \sim \epsilon \log(\epsilon \ell) + \sum_{n=1}^\infty \frac{\epsilon^{n+1}}{\ell^n} ~, \quad {}_s \delta_{\ell m}^{\rm NZ} \sim \epsilon^{2\ell + 1} \sim (r_s \omega)^{2\ell + 1} ~.
\end{equation}
In this regime, except for the logarithmic dependence $\log(\ell)$, the far zone shows the power law decay while the near zone features an exponential decay when $\ell \rightarrow \infty$. Physically, the logarithmic term reveals the long-range nature of the Newtonian $G M/r$ potential. The power law behavior $1/\ell^n$ indicates that all the  PM-corrections share the non-local power law decay in the potential $(G M / r)^n,n \geq 2, n \in \mathbb{N}$ when the radius $r \rightarrow \infty$. The near zone phase-shift reveals the common feature for the  scattering against  a localized potential, i.e. potentials with exponential decay also known as  ``hard-sphere" scattering \cite{Landau:1991wop}, where the low-energy scattering at large $\ell$ shares the universal behavior $(\omega R)^{2\ell+1}$, with  $R$ the range of the potential.

\section{Far-zone, tree-level gravitational Compton Amplitude  for Kerr}
In this section we analyze the point-particle limit of massless perturbations of the KBH in the context of the tree-level,  helicity preserving, Gravitational Compton amplitude. As discussed above, this limit  can be      studied completely from the  far zone  contributions to the phase-shift \eqref{eq:phase-far} while ignoring the near zone tidal effects capture by \eqref{eq:phase-near}. We recall we use the generic-$\ell$ prescription. 

Consider then a $s=-2$ plane wave scattering off KBH. The wave  impinges with momentum $k_2^\mu$ and scatters with  momentum $k_3^\mu$, over the BH with momentum $p_1^\mu=M u^\mu$. The angle formed by the direction of the impinging wave  and the BH's spin, $a_{\text{BH}}^\mu$, is $\gamma$. The far zone, helicity preserving amplitude is computed from the infinite sum of harmonics  
\cite{Dolan:2008kf}
\begin{equation}
\mathcal{A} {=}\sum_{\ell,m}\left[{}_{-2}S_{\ell m}\left(\gamma,0;\frac{\epsilon \chi}{2}\right){}_{-2}S_{\ell m}\left(\vartheta,\varphi;\frac{\epsilon \chi}{2}\right) \mathcal{ A}_{\ell,m} \right]\,,\label{Eq:fKerrGeneric}
\end{equation}
where 
\begin{equation}
\label{eq:amplitude_modes}
\mathcal{A}_{\ell m} =\frac{2\pi}{i\omega} \sum_{P=\pm1} \left(e^{{}_{-2}\delta^{P,\text{FZ}}_{\ell m}(2i)}-1\right)\,.
\end{equation}
The sum over   $P$ is due to the  change from the parity to helicity basis. 
As mentioned, we are  interested in the  contribution to the phase-shift of the form $\chi^i\epsilon^{i+1}\sim GM\omega (a_{\text{BH}}\omega)^i$; which exhibit explicitly a tree-level scaling.  
From the  analysis based on the properties of the NS-function presented above, we know that  ${}_{-2}\delta^{P,\text{FZ}}_{\ell m}$   not only captures  purely conservative effects, but it is merely a polynomial function in $\chi$, making straightforward to extract its tree-level contributions to the amplitude. In fact, because of this polynomiality, these contributions are exact for Kerr in the sense that no analytic continuation in $\chi\gg1$ is required. 

Since we are using the generic$-\ell$ prescription, at this point one might worry that the $\ell$-poles discussed around eq. \eqref{eq: generic ell} will be problematic since we are dropping the near zone contributions responsible to cancel them. Fortunately, this poles do not contribute to the tree-level amplitude and  can be safely ignored \footnote{In the partial wave basis, the $\ell$-poles show up when the loop diagram has UV divergence \cite{loop}.}. In addition, since the $\epsilon$-corrections in $a$  are of the form $ \sum_i c_i \chi^i \epsilon^{i+(n\ge2)} $,  we can simply use the tree-level value $a=-\frac{1}{2}-\ell$; this in turn provides a great simplification in the computation of ${}_{-2}\delta^{P,\text{FZ}}_{\ell m}$.

At leading order in the PM-expansion, we can arrange the amplitude modes
\eqref{eq:amplitude_modes} as
\begin{equation}\label{Eq:phasetobeta}
 \mathcal{A}_{\ell m}{=} 4\pi GM e^{i\Phi} \left( \psi ^{(0)}(\ell{-}1)+\psi ^{(0)}(\ell{+}3){+}\sum_{i=1}
    \beta_{\ell m} ^{(i)}(\chi\epsilon)^{i}  \right)\,,
 \end{equation}
where $\Phi=2\epsilon\log(2\epsilon)-\epsilon\to 0$, and  we have written  the explicit tree-level combinations $(\chi \epsilon)^{i}$, allowing for the $\beta$-coefficients\footnote{These receive  contributions only from $\frac{1}{2} {\rm Im} \big[\partial_{m_3} F \big] {+} \frac{ \kappa}{2} \epsilon $ 
in 
eq. \eqref{eq:phase-far}. } to be  functions only of $\ell$ and $m$.  These coefficients are  real, which makes manifest   the conservative character of \eqref{Eq:phasetobeta}.
 We include the  $\beta_{\ell m}^{(i)}$ coefficients needed to compute the tree-level far zone amplitude modes \eqref{Eq:phasetobeta},  up to $\mathcal{O}(a_{\text{BH}}^8)$
in the ancillary files associated to the \texttt{arXiv} submission of this work \cite{ancill}.

Following \cite{Bautista:2021wfy,Bautista:2022wjf}, we match \footnote{This match is to be done by expanding the spheroidal harmonics in \eqref{Eq:fKerrGeneric}, in a basis of spin-weighted spherical harmonics \cite{Dolan:2008kf}. Since the details of this matching have been widely discussed in \cite{Bautista:2021wfy,Bautista:2022wjf}, we shall not provide them here.   }   amplitude \eqref{Eq:fKerrGeneric} together with  \eqref{Eq:phasetobeta}, to a covariant tree-level classical ansatz of the form\footnote{In this work we used the all-orders in spin ansatz given by eq. (3.47) in \cite{Bautista:2022wjf}. Additional Ansätze  have been considered previously \cite{Bern:2022kto,Aoude:2022trd,Aoude:2022thd,Chiodaroli:2021eug,Haddad:2023ylx}, including   those from    higher-spin gauge theory \cite{Cangemi:2022bew,Cangemi:2023ysz}. We thank the authors of the last two references  for sharing their unpublished  $\mathcal{O}(a_{BH}^8)$ gravitational ansatz with us. 
} 
\begin{equation}\label{eq:ansatzspin}
  \mathcal{A} =  \mathcal{A}^{(0)} \left(e^{w-x} + f_\xi^{\text{FZ}}(x,y,w) \right )\,,
\end{equation}
where $\mathcal{A}^{(0)}$ is the tree-level gravitational Compton amplitude for Schwarzschild BH and $\xi=(u\cdot(k_2+k_3))^2/(k_3-k_2)^2$, is the scattering optical parameter.  The contact term function, $f_\xi^{\text{FZ}}(x,y,w)$,  is chosen in such way, the spurious pole  in $w^{\ge5}$, from expanding the exponential function in \eqref{eq:ansatzspin}, is canceled. 
Here we have used the conventions of  \cite{Cangemi:2023ysz} to write the spin operators:  $x= (k_3{+}k_2){\cdot} a_{\text{BH}},\,y= (k_3{-}k_2){\cdot} a_{\text{BH}},\, w= \frac{2u\cdot k_2}{u\cdot \epsilon_2} \epsilon_2{\cdot} a_{\text{BH}}$, and the  gauge 
$
\epsilon_2^\mu=\frac{ \langle 2|\sigma^\mu|3]}{\sqrt{2}[32]} \propto \tilde{\epsilon}_3^\mu = \frac{ \langle 2|\sigma^\mu|3]}{\sqrt{2}\langle32\rangle }\,$, with $|i\rangle, |i]$ the  spinors of the massless momentum $k_i$.

After uniquely matching     the modes of  \eqref{Eq:phasetobeta}  to those obtained from \eqref{eq:ansatzspin}, 
  we finally obtain the contact terms    $f_\xi^{\text{FZ}}(x,y,w)$, entering \eqref{eq:ansatzspin} to be:
\begin{widetext}
\begin{equation}\label{eq:ComptonFinal}
\begin{split}
 f_\xi^{\text{FZ}}(x,y,w)=&\frac{1}{8!\times 9}\left[ -128 \left(w^7 (2 w+3 x-6)\right) \xi ^{-2}+64 \upsilon _1 w^5 \xi^{-1}-8 \upsilon _2 w^3 (w-x-y) (w-x+y)\right.\\
&\left.\qquad-4  \upsilon _3 w \left((w-x)^2-y^2\right)^2 \xi - \upsilon _4 \left((w-x)^2-y^2\right)^3\xi ^2 -278  \left((w-x)^2-y^2\right)^4\xi ^3\right]+\mathcal{O}(a_{\text{BH}}^9)\,,
\end{split}
\end{equation}
where
\begin{align}
    \upsilon _1=&-59 w^3+3 w^2 (72-43 x)+w \left(5 x (60-23 x)+y^2-504\right)+3 x \left(x (68-19 x)+y^2-168\right)-6 y^2+756\,,\\
     \upsilon_2=& \,1199 w^3 {+} 33 w^2 ( 77 x{-}114 ) 
 {+} 
 3 w (2520 {+} x (  631 x{-}1668) {+} 23 y^2){+} 
 3 x (1680 {-} 638 x {+} 169 x^2) {+} 33 (x{-}2) y^2  {-}7560\,,\\
 \upsilon _3=&\,2285 w^3{+}w^2 (3871 x{-}5484){+}w \left(x (1949 x{-}5172){+}97 y^2{+}7560\right){+}9 \left(x \left(x (27 x{-}104){+}3 y^2{+}280\right){-}6 \left(y^2{+}70\right)\right)\,,\\
 \upsilon _4=&\,\, 3355 w^2+6 w (581 x-804)+x (659 x-1752)+37 y^2+2520\,.
\end{align}
\end{widetext}

Up to  $\mathcal{O}(a_{\text{BH}}^6)$, amplitude \eqref{eq:ComptonFinal} agrees with the results reported in Table 1 in \cite{Bautista:2022wjf} for $\alpha=1$ and $\eta=0$, computed from the MST-method using the fixed-$\ell$ prescription\footnote{For recent uses of Compton amplitude in the two-body context see \cite{Guevara:2018wpp,Chung:2018kqs,Bautista:2019tdr,Bautista:2021inx,Chen:2021kxt, 
Bern:2022kto,Aoude:2022trd,Aoude:2022thd,Chiodaroli:2021eug,Haddad:2023ylx,Cangemi:2022bew,Bjerrum-Bohr:2023jau,Alessio:2023kgf,DeAngelis:2023lvf,Aoude:2023dui,Brandhuber:2023hhl}.}. As expected, the far zone amplitude is independent of  horizon effects and the boundary conditions  used at $r_+$. Notice then that while in the fixed-$\ell$ prescriptions used in  \cite{Bautista:2022wjf}, the super extremal (SE) limit, $\chi\gg1$, was needed to disentangle near from far physics effects, in the generic-$\ell$ prescription this continuation is not needed. In fact, as one can show, using the latter prescription, and after removing dissipative contributions,  the dropped near-zone pieces vanish in the SE-limit (see Appendix A for a specific example). 
In addition,
terms tagged with $\alpha$ in the results reported in \cite{Bautista:2022wjf}
were interpreted as contributions  coming from digamma functions in the SE-limit\footnote{The $\alpha$-tags were added after identity \eqref{eq:identitypoly}, relating special polygamma functions, was used.}.   Their appearance in the point-particle amplitude is  just an artefact of using the fixed-$\ell$ prescription for computing the total phase-shift, which mixes the near and far zone effects,  as explained above.  From the discussion here, we conclude then no polygamma contributions actually appear in the point-particle Compton amplitude. 

\textit{Helicity reversing amplitude}: In an analogous computation, we have checked that the 
  helicity reversing   gravitational Compton amplitude  extracted \textit{purely} from the far zone contribution to the phase-shift \eqref{eq:phase-shift-gen}, agrees with the minimal coupling  exponential $\sim e^{y}$ up to eight order in spin.

\section{Discussion} 
In this work we have shown  a new perspective on black hole perturbation theory computations based on the use of NS-functions which makes manifest analytic properties and symmetries otherwise obscured by using other methods. It is desirable to further study the NS-function aiming to provide additional non-perturbative analytic results in classical physics. 

Along these lines, in this work we have shown a natural separation between near and far zone physics, based on a generic-$\ell$ prescription. 
As it is well known \cite{Glampedakis:2001cx,Bautista:2021wfy}, this prescription is  powerful for estimating the eikonal limit -- $\ell \rightarrow \infty, \omega \rightarrow \infty$ while $\omega/ \ell$ is fixed -- of the classical observables.  In this limit,  results  are universal  --independent of the spin-weight $s$ of the perturbation -- and receive contributions  purely from far zone-physics. Studying the NS-function in this limit and its connection to geodesic motion is left for future work.

It is interesting to note also how the near-far factorization provides a natural separation of the spectrum of the theory. As it is well know, it can be access through the poles of the scattering amplitude.  From \eqref{eq:nearpfarCFT}, we thus identify two distinct types of poles in the Compton  amplitudes. Firstly, in the eikonal limit -- which only involves the far zone contributions --  and at leading order in $\epsilon$, the infinite sums of harmonics produce gamma functions whose poles  correspond to the bound states of the Newtonian potential, whose locations are $\epsilon=i\,,2i\,,3i\,,\cdots$ \cite{Bautista:2021wfy}. 
The second type of poles come from near zone physics; they correspond to the quasi normal mode (QNM) resonances, for which the exact quantization condition follows as \cite{Aminov:2020yma,Bonelli:2021uvf} 
\begin{equation}
    1 + e^{i\pi a} \frac{\cos(\pi(m_3 - a))}{\cos(\pi(m_3 + a))} \mathcal{K} =0 ~.
\end{equation}
This relation establishes a direct link between the tidal response function $\mathcal{K}$ and the QNM spectrum. Interestingly, we also find that $\mathcal{K}$   can be written in terms of the full (instanton plus one-loop) NS free energy  $F_{\rm full}$ \cite{Aminov:2020yma,Bonelli:2021uvf}
\begin{equation}
    e^{i\pi a} \frac{\cos(\pi(m_3 - a))}{\cos(\pi(m_3 + a))} \mathcal{K} = - e^{\partial_a F_{\rm full}} ~,
\end{equation}
which could lead to some hidden structures to be further investigated. 
Additional  methods based on the Thermodynamic Bethe Ansatz 
have been used to study applications of NS functions to QNMs \cite{Fioravanti:2021dce,Fioravanti:2022bqf} and it would be interesting to study them in the context of scattering amplitudes.

Away from the point particle limit, Compton amplitudes receive contributions from the 
 near-zone
phase-shift \eqref{eq:phase-near} starting at order $\epsilon^{2\ell+1}$ -- the order at which the BH horizon effects become important. -- Intriguing, at this order the phase-shift   comes with special polygamma functions of the form $\psi ^{(n)}(i\frac{m \chi }{\kappa} {\pm} \ell)$.  Inspection  of near-zone phase-shift \eqref{eq:phase-near} suggests  that when $\chi\le1$ the near-zone piece
 does not provide any tree-level information (see also Appendix A for a explicit example). 
There is however a subtlety with  this observation since from the 
 \begin{align}\label{eq:identitypoly}
  \psi ^{(n)}(z{\pm} \ell){=} \psi ^{(n)}(z)\pm \sum_{k{=}0{+}\eta_\pm}^{\ell{-}1{+}\eta_\pm} \frac{({-}1)^n n!}{(z\pm k)^{n+1}}\,,
\end{align}  
for  $n,\ell\in \mathbb{Z}^+ $ and $(\eta_+,\eta_-)=(0,1)$, polynomial contribution with tree-level scaling  arise from \eqref{eq:phase-near}, by means of the  sum term in \eqref{eq:identitypoly}.   It is therefore ambiguous to extract tree-level contributions from the near-zone phase-shift without invoking an analytic continuation in the Kerr spin parameter, since the association of polygamma contributions to loop effects can be done either before or after identity \eqref{eq:identitypoly} has been used. 
Interestingly, since  the additional tree-level contributions arising from the sum term in \eqref{eq:identitypoly}
can appear  only once the square-root from $\kappa=\sqrt{1-\chi^2}$ is removed, i.e. from  $\kappa^{2n}$ terms, and since  $\kappa$ inside the polygamma functions \textit{always} comes accompanied by a factor of $i$ (see also \eqref{eq:dict}), then tree-level scaling implies that factors involving  $(i\kappa)^{2n}$ are  purely  real,  signaling absorptive effects in a 4-point amplitude \textit{can never} come in tree-level form. In an on-shell language, matching of absorptive effects  to effective tree-level-like three-point has been consider  recently \cite{Aoude:2023fdm,Jones:2023ugm,Chen:2023qzo}.
The inclusion/interpretation of near-zone effects,  their interplay with the constraints from  dynamical   multipole-moment  on the Gravitational Compton amplitude recently proposed in \cite{Scheopner:2023rzp}
and the translations of the constrains imposed by the  symmetry  of the NS-function to the Compton amplitudes written in the  covariant basis are left for future investigation.

Finally, the technology used   here in the context of linear  perturbation theory  can be naturally imported to the  study of  non-linear perturbations of KBHs, since,  the second order Teukolsky equations are still of  confluent Heun-type  but with the addition of  nonlinear source, obtained  from the first order solution \cite{Loutrel:2020wbw,Ripley:2020xby}. A comprehensive analysis utilizing this novel method and their interplay with second order self force approaches are left future investigation.

\sectionskip
\Section{Acknowledgments.}   
We would like to thank Paolo Arnaudo, Alba Grassi, Aidan Herderschee, Misha Ivanov, Henrik Johansson, 
Chris Kavanagh, Yue-Zhou Li, Francisco Morales,  Julio Parra-Martinez, Giorgio Di Russo and  Justin Vines 
for useful discussions and comments on an early draft of this paper. We specially thank Chris Kavanagh  for pointing out  identity \eqref{eq:identitypoly}, and  for sharing his fixed-$\ell$  MST results for  SE-Kerr Compton amplitude up to order $a_{\text{BH}}^7$, which are in complete agreement with \eqref{eq:ComptonFinal} up to this order. This work makes use of the Black Hole Perturbation Toolkit \cite{BHPToolkit}.
The work of Y.F.B.   has been supported by the European Research Council under Advanced Investigator Grant ERC–AdG–885414. The work of C.I. is partially supported by the Swiss National Science Foundation Grant No. 185723.
The research of G.B. is partly supported by the INFN Iniziativa Specifica ST\&FI and by the PRIN project “Non-perturbative Aspects Of Gauge Theories And Strings”. The research of  A.T. is partly supported by the INFN Iniziativa Specifica GAST and InDAM GNFM. 
The research of G.B and A.T. is partly supported by the MIUR PRIN Grant 2020KR4KN2 "String Theory as a bridge between Gauge Theories and Quantum Gravity".  
G.B. and A.T. acknowledge funding from the EU project Caligola HORIZON-MSCA-2021-SE-01), Project ID: 101086123.

\bibliography{references}

\newpage
\onecolumngrid
\newpage

\appendix

\section{Appendix A: MST Method Review and Near-Far Factorization}
\label{app_A}

We start this  appendix by reviewing the  MST method 
~\cite{Mano:1996mf,Mano:1996vt,Mano:1996gn,Sasaki:2003xr} for solving  TME using matching asymptotic expansion, where the  renormalized angular momentum $\nu$ is introduced. As a consequence of the matching asymptotic expansion, we then  discuss the near-far factorization in the Compton scattering phase-shift ${}_s\delta_{\ell m}$. Finally, we explicitly show that in the generic $\ell$ prescription, there are spurious poles in the MST coefficients when $ \ell \in \mathbb{N}$ and it will be cancelled when adding near-zone and far zone 

\subsection{MST Method Review }
In the MST approach, one first constructs the near-zone solution based on a double-sided infinite series of hypergeometric functions which converges within $r_+ \leq |r| < \infty$ \footnote{Here, the radial coordinate $r$ takes the value in $\mathbb{C}_\infty$.}. This convergence radius gives a natural definition for the near-zone:
\begin{equation}
    \textbf{MST Near Zone}: r_+ \leq |r| < \infty ~.
\end{equation}
Similarly, one then constructs the far zone solution based a double-sided infinite series of Coulomb wavefunction which converges within $r_+ < |r| \leq \infty$, which can be used as the definition of far zone:
\begin{equation}
    \textbf{MST Far Zone}: r_+ < |r| \leq \infty ~.
\end{equation}

\begin{figure}[h!]
    \centering
    \includegraphics[scale = 0.15]{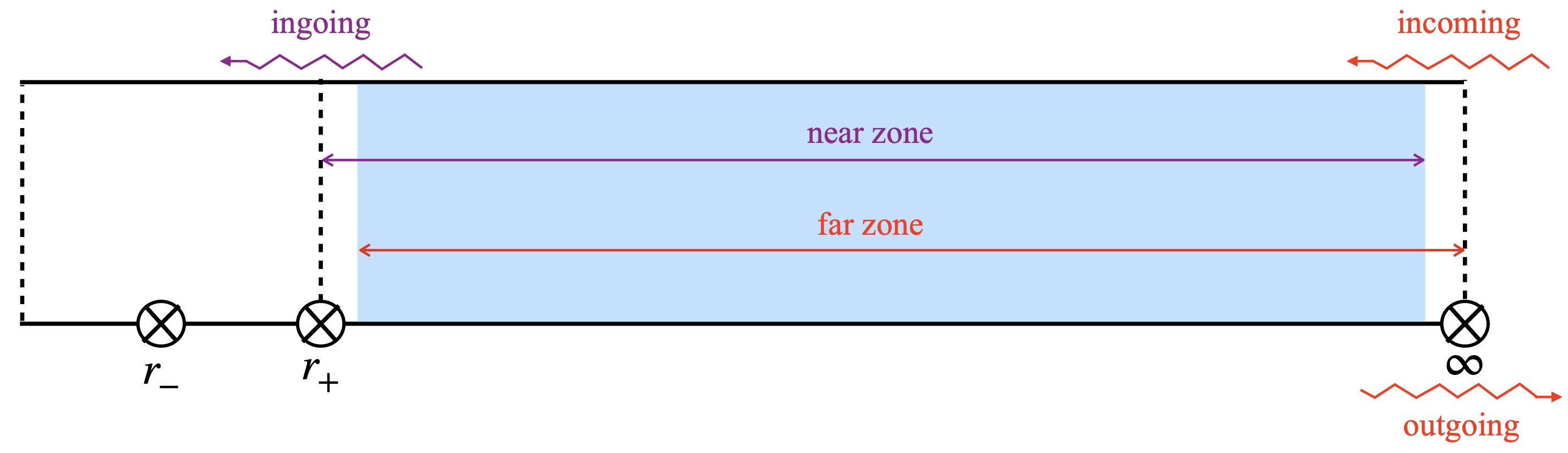}
    \caption{A schematic diagram illustrates the convergence radius of near and far zone solutions. The blue region denotes the overlapping region where the matching is performed.}
    \label{fig:matching}
\end{figure}

To get the solution that is converging everywhere, one needs to match the near-zone solution with far-zone solution in the overlapping region $r_+ < r<\infty$. In \cref{fig:matching}, we show a schematic diagram for the matching asymptotic expansion. 
To ensure the convergence and the matching of the solutions on both sides, one needs to introduce an auxiliary non-integer parameter, the so-called renormalized angular momentum $\nu(s,\ell,m, \omega)$, which is a function of spin-weight $s$, angular momentum $\ell$, azimuthal quantum number $m$, and the frequency $\omega$ of the perturbation. In the low-frequency region it has the form
\begin{equation}
\label{eq: low-frequency nu}
    \nu=\ell+\frac{1}{2 \ell+1}\left(-2-\frac{s^2}{\ell(\ell+1)}+\frac{\left[(\ell+1)^2-s^2\right]^2}{(2 \ell+1)(2 \ell+2)(2 \ell+3)}-\frac{\left(\ell^2-s^2\right)^2}{(2 \ell-1) 2 \ell(2 \ell+1)}\right) \epsilon^2+\mathcal{O}\left(\epsilon^3\right) ~,
\end{equation}
where $\epsilon \equiv 2 G M \omega$. Formally, $\nu$ is known as the characteristic exponent because it governs the following asymptotic behavior of the near-zone ingoing solution (see Eq.~(166) in \cite{Sasaki:2003xr})
\begin{equation}
    {}_s R_{\ell m}^{\rm near} \sim K_{\nu} r^{\nu} + K_{-\nu-1} r^{-\nu-1} ~, \quad r \rightarrow \infty ~.
\end{equation}
The coefficient ratio $K_{-\nu-1}/K_{\nu}$ tells the relative amplitude between the decaying $r^{-\nu-1}$ and the growing $r^\nu$ mode which captures the BH tidal response. As mentioned in the main text, a detailed comparison between CFT method and MST method shows that $\mathcal{K}$ given in \eqref{eq:nearpfarCFT} agrees with $K_{-\nu-1}/K_{\nu}$ and thus we call $\mathcal{K}$ the BH tidal response function. After performing the matching, and imposing the appropriate boundary conditions \eqref{eq:bdary},
one finally obtains the wave amplitude ratio as follows:\footnote{see Eq.(168) and Eq.(169) in Ref.~\cite{Sasaki:2003xr} for more explicit expressions and Eq.~(12) in \cite{Ivanov:2022qqt} for the first proposal of the factorized form.}
\begin{equation}\label{eq:factorization}
\frac{B_{s \ell m}^{\text{refl}}}{B_{s \ell m}^{\text{inc}}}{=}\frac{1}{\omega^{2s}}{\blue{\underbrace{\frac{1+ie^{i\pi\nu}\frac{K_{-\nu-1;s}}{K_{\nu;s}}}{1{-}ie^{-i\pi\nu}\frac{\sin(\pi(\nu{-}s{+}i\epsilon))}{\sin(\pi(\nu+s-i\epsilon))}\frac{K_{{-}\nu{-}1;s}}{K_{\nu;s}}}}_{\text{near\,zone}}}}{\red{\underbrace{\times\frac{A_{-;s}^{\nu}}{A_{+;s}^{\nu}}e^{i\epsilon(2\ln\epsilon-(1-\kappa))}}_{\text{far\,zone}}}}\,,
\end{equation}
where
\begin{equation}\label{eq:farz}
    \frac{A_{-;s}^\nu}{A_{+;s}^\nu} =\frac{e^{-{\pi}i(\nu+1)}}{ 2^{2(s-i\epsilon)}} \times \frac{\Gamma(\nu + 1 + s -i \epsilon)}{\Gamma(\nu + 1 -s + i \epsilon)} \times \frac{
\sum_{n=-\infty}^{+\infty}(-1)^n{(\nu+1+s-i\epsilon)_n\over 
(\nu+1-s+i\epsilon)_n}a_n^\nu}{\sum_{n=-\infty}^{+\infty}a_n^\nu}\,,
\end{equation}
and
\begin{equation}\label{eq:ks}
    \frac{K_{-\nu-1;s}}{K_{\nu;s}} = (2 \epsilon \kappa)^{2\nu+1} \frac{\Gamma (-2 \nu -1) \Gamma (-2 \nu ) \Gamma (\nu -i \tau +1) \Gamma (-s-i \epsilon +\nu +1) \Gamma (-s+i \epsilon +\nu +1)}{\Gamma (2 \nu +1) \Gamma (2 \nu +2) \Gamma (-\nu
   -i \tau ) \Gamma (-s-i \epsilon -\nu ) \Gamma (-s+i \epsilon -\nu )} \times \frac{X_{-\nu-1}}{X_{\nu}} ~.
\end{equation}
Here, $ \kappa = \sqrt{1 - \chi^2}$ and $\tau = (\epsilon - m \chi)/\kappa$.
The function $X_{\nu}$ is given by the product of two infinity sum of the MST coefficients,  $a_n, -\infty\leq n \leq +\infty$;
\begin{equation}\label{eq:X-MST}
\begin{aligned}
X_\nu &= \left(\sum_{n=0}^{+\infty} \frac{(-1)^n}{n!} (1+2\nu)_n a_n^\nu \frac{(1+\nu+s+i\epsilon)_n}{(1+\nu-s-i\epsilon)_n}\frac{(1+\nu+i\tau)_n}{(1+\nu-i\tau)_n}\right) \times \left(\sum_{n=-\infty}^{0}
	\frac{(-1)^n}{(-n)!
	(2\nu+2)_n}\frac{(\nu+1+s-i\epsilon)_n}{(\nu+1-s+i\epsilon)_n}
	a_n^{\nu}\right)^{-1} \, ~.
\end{aligned}
\end{equation}
 MST coefficients  satisfy the following three term recurrence relation along with the ``renormalized" angular momentum $\nu$
\begin{equation}\label{eq: recurrence relation}
    \alpha_n^\nu a_{n+1}^\nu + \beta_n^\nu a_{n}^\nu + \gamma_n^\nu a_{n-1}^\nu = 0 ~,
\end{equation}
where
\begin{equation}\label{eq: recurrence parameter}
    \begin{aligned}
\alpha_n^\nu & =\frac{i \epsilon \kappa(n+\nu+1+s+i \epsilon)(n+\nu+1+s-i \epsilon)(n+\nu+1+i \tau)}{(n+\nu+1)(2 n+2 \nu+3)} \\
\beta_n^\nu & =-{}_s \lambda_\ell^m -s(s+1)+(n+\nu)(n+\nu+1)+\epsilon^2+\epsilon(\epsilon-m \chi)+\frac{\epsilon(\epsilon-m \chi)\left(s^2+\epsilon^2\right)}{(n+\nu)(n+\nu+1)} \\
\gamma_n^\nu & =-\frac{i \epsilon \kappa(n+\nu-s+i \epsilon)(n+\nu-s-i \epsilon)(n+\nu-i \tau)}{(n+\nu)(2 n+2 \nu-1)} .
\end{aligned}
\end{equation}
We fix $a_0=1$ for convenience.
These recurrence relations can be solved order by order in the PM expansion.

Let us finally provide the explicit expressions for the Teukolsky-Starobinsky constant $A_s^P$ entering \eqref{eq:phase-shift-gen}
\begin{equation}
\label{eq:teukstarovins}
    \begin{aligned}
       & A_0^{\pm} =  1  \,,\\
       & A_{-1}^{\pm} = \left[\left( {}_{-1}Q_{\ell m}  +a^2 \omega^2-2 a m \omega\right)^2+4 a m \omega-4 a^2 \omega^2\right]^{1 / 2} ~, \\
       & A_{-2}^{\pm} = \Bigg[ \left( ({}_{-2} Q_{\ell m})^2+4 a \omega m-4 a^2 \omega^2\right)\left[({}_{-2} Q_{\ell m} - 2)^2+36 a \omega m-36 a^2 \omega^2\right] \\
        & \quad +(2 {}_{-2}Q_{\ell m} - 1)\left(96 a^2 \omega^2-48 a \omega m\right)-144 \omega^2 a^2 \Bigg]^{1/2} \pm i 12 M \omega ~,
    \end{aligned}
\end{equation}
where ${}_s Q_{\ell m} \equiv {}_s \lambda_{\ell m} + s (s+1)$. ${}_s\lambda_{\ell m}$ is the angular eigenvalue of the spheroidal harmonics.

\subsection{Near-Far Factorization}

The near-far factorization proposed in \cite{Ivanov:2022qqt, Saketh:2023bul} shows that the Kerr Compton scattering phase shift once expanded in the small frequency limit, i.e. $G M \omega \ll 1$ can be directly separated into the near zone and far zone contributions. The far-zone phase shift has the following feature
\begin{equation}
    {}_s \delta_{\ell m}^{\rm FZ} \sim (G M \omega) \log(G M \omega) + (G M \omega) + (G M \omega)^2 + (GM \omega)^3 + \cdots  
\end{equation}
which features integer power of $G$ scaling except for the logarithmic term due to the scattering off long-range Newtonian potential. Higher order in $G$ corrections can be understood as the PM corrections upon the point-particle approximation. The near-zone phase shift features non-analytic behavior of $G$
\begin{equation}
    {}_s \delta_{\ell m}^{\rm NZ} \sim (G M \omega)^{2\nu + 1} (1 + (G M \omega) + (G  M \omega)^2 + \cdots) ~,
\end{equation}
for generic value of $\nu$. Once performing the low-frequency expansion, the non-analyticity leads to the logarithmic corrections
\begin{equation}
    \left.(G M \omega)^{2 \nu+1}\right|_{\nu=\ell+\mathcal{O}\left((G M \omega)^2\right)+\ldots}=(G M \omega)^{2 \ell+1}\left(1+(G M \omega)^2 \log (G M \omega)+\cdots\right) ~,
\end{equation}
which have a natural  understanding in terms of  the renormalization group (RG), where are   running of ``dynamical" Love numbers for Kerr BHs appears \cite{Saketh:2023bul}.

\subsection{phase-shifts for $s=0,\ell=1,m=1$ Perturbations}
For illustrative purposes, let us close this appendix by  explicitly showing   the cancellation of the $\ell$-poles at the level of the phase-shift in the  scalar example presented in the main text. We keep up to $\mathcal{O}(\epsilon^4)$, which for the $s=0,\ell=m=1$ case, is the order at which the first poles appear.
In the generic-$\ell$ prescription, the near- and far- zone contributions take the form respectively
\begin{equation}
\label{eq:nears0l1m1}
\begin{split}
{}_0 \delta_{\ell 1}^{\rm NZ}|_{\ell {\rightarrow} 1} {=}&  \blue{ \( \frac{\chi }{72 (\ell {-}1)}{+} \(\frac{\gamma_E}{18} - \frac{7}{54} \)\chi + \frac{1}{36} \chi \log(2 \epsilon \kappa) + \frac{1}{36} \chi {\rm Re}\[\psi^{(0)}\(\frac{i \chi}{\kappa} -1\) \]\) \epsilon^4 }
 \end{split}
\end{equation}
 and 
\begin{equation}
\label{eq:fars0l1m1}
\begin{split}
{}_0\delta_{\ell 1}^{\rm FZ}|_{\ell {\rightarrow} 1} & = \red{\epsilon \log (2 |\epsilon|) + \(-\frac{3}{2} + \gamma_E\) \epsilon + \(\frac{19}{60} \pi - \frac{\chi}{4}\) \epsilon^2+ \(\frac{19}{180} \pi^2 - \frac{7 \pi \chi}{60} + \frac{\chi^2 }{40} - \frac{\zeta(3)}{3}\) \epsilon^3} \\
& \quad + \red{\( - \frac{\chi}{72(\ell - 1)} + \frac{78037}{378000} \pi - \frac{130 + 42\pi^2}{1080} \chi + \frac{143}{4200} \pi \chi^2 + \frac{\chi^3}{40} \) \epsilon^4} ~.
\end{split}
\end{equation}
The same computation can be done in the 
 fixed-$\ell$ prescription. Using the MST coefficients listed in Appendix B in \cite{Casals:2016soq},   we have\footnote{As noted in footnote:\footref{fn:PMMST}, whereas in the generic $\ell$ prescription, the MST coefficients $a^\nu_{\pm n}$ scale symmetrically in $\epsilon$: that is $a^\nu_{\pm n}\sim \epsilon^{|n|}$, in the fixed-$\ell$ prescription this symmetric scaling is lost.  }
\begin{equation}
\label{eq:nearfars0l1m1}
\begin{split}
{}_0\delta_{11}=&
\red{ \epsilon \log(2|\epsilon|) + \left({-}\frac{3}{2}{+}\gamma_E \right) \epsilon + \left( \frac{19 \pi }{60}{-}\frac{13 \chi }{57}\right) \epsilon ^2 + \left(\frac{19 \pi ^2}{180}{-}\frac{1583 \chi ^2}{137180}{-}\frac{9 \pi  \chi }{95}{-}\frac{\zeta (3)}{3}{-}\frac{5}{228}\right) \epsilon ^3} \\
&\red{ + \left(\frac{8100833 \chi ^3}{831969264}{-}\frac{17947 \pi  \chi ^2}{7201950}{-}\left(\frac{92867}{2880780}{+}\frac{3 \pi ^2}{95}\right) \chi {+}\frac{1325203 \pi }{7182000}\right) \epsilon ^4} \\
& 
\blue{{-}\frac{5 \chi  \epsilon ^2}{228}{+}\left( \frac{5}{228} {-} \frac{5 \pi }{228} \chi + \frac{2005 \chi ^2}{54872} \right) \epsilon ^3{+}\epsilon ^4 \left( \frac{5 \pi }{228} + \frac{2005 \pi }{54872}  \chi ^2 +\frac{63491993 }{4159846320} \chi ^3 \right.}\\
&\blue{\left.
+\chi  \left(\frac{1}{36} {\rm Re}\[ \psi ^{(0)}\left(\frac{i \chi }{\kappa }-1\right) \] +\frac{1}{36} \log (2 \kappa  \epsilon )-\frac{5\pi^2}{684}-\frac{156832}{720195}+\frac{\gamma_E}{18}\right)\right)}\,.
\end{split}
\end{equation}
By combining \eqref{eq:nears0l1m1} and \eqref{eq:fars0l1m1}, we find that the diverging terms $1/(\ell - 1)$ cancel, aligning perfectly with the results shown in \eqref{eq:nearfars0l1m1}.  The colors in the fixed-$\ell$ results denote a hypothetical near-far factorization. From \eqref{eq:nearfars0l1m1}, we see that with the fixed-$\ell$ prescription, there are no singular contributions  in any region, but this comes at the cost of mixing  the terms from the near and far zones.
For instance, the tree-level contributions, i.e. terms scale as $\chi^i\epsilon^{i+1}$, entirely come from the far zone in generic-$\ell$ prescriptions
\begin{equation}\label{eq:tree-scal}
    \red{\underbrace{\frac{1}{40} \chi^3 \epsilon ^4+\frac{\chi^2 \epsilon ^3}{40}-\frac{\chi \epsilon ^2}{4}}_{{\rm generic}-\ell
    \,{\rm far\, zone}}} = \red{\underbrace{-\frac{13\chi}{57} \epsilon^2 - \frac{1583\chi^2}{137180} \epsilon^3 + \frac{8100833\chi^3}{831969264} \epsilon^4 }_{{\rm fixed}-\ell \,{\rm far\, zone}}} \blue{\underbrace{-\frac{5\chi}{228} \epsilon^2 + \frac{2005\chi^2}{54872} \epsilon^3 + \frac{63491993 \chi^3}{4159846320} \epsilon^4}_{{\rm fixed}-\ell \,{\rm near \, zone}}}~,
\end{equation}
while the fixed-$\ell$ prescription splits these terms into unusual and confusing mixes.

 There is however a subtlety when extracting tree-level contributions
as showed in the discussion section. From  \eqref{eq:nears0l1m1} no apparent tree-level contribution arises for $\chi\le1$, however if  identity \eqref{eq:identitypoly} was used, the tree-level,  contribution $-\chi^3\epsilon^4/36$ will be extracted from \eqref{eq:nears0l1m1}. 
 To avoid this subtlety, and  in order to 
extract tree-level contribution in the point particle limit, in Refs. \cite{Bautista:2021wfy,Bautista:2022wjf} the super-extremal (SE) limit was necessary. However, because of the near-far zone mixing in the fixed-$\ell$ prescription used in those references, in combination of the use of identity \eqref{eq:identitypoly}, an apparent contribution from digamma function, tagged with the $\alpha$-label, appeared in the point-particle amplitude. The $\alpha$-label was added to the digamma appearing in the right-hand-side of \eqref{eq:identitypoly}. 
In the generic-$\ell$ prescription, this mix does not take place and, as one can explicitly check, no tree-level contribution arises from \eqref{eq:nears0l1m1} in the SE-limit.  

The resulting covariant tree-level amplitude computed with \eqref{eq:tree-scal} 
agrees    with the results in eq. (4.54-4.55) in \cite{Bautista:2021wfy}. The  extra $\chi^3\epsilon^4/36$ would change such results    precisely canceling the   contact terms modifying the Born amplitudes eq. (4.54-4.55) in \cite{Bautista:2021wfy}.

It is also interesting to analyze the contribution from the dissipative pieces. The absorption probability $\Gamma \sim \sum_{P} [1- ({}_s\eta_{\ell m}^{P})^2]$ can be estimated from
\begin{equation}
\begin{split}
{}_0 \eta_{1 1}^P {=}
\blue{1 +
\frac{\epsilon ^3}{36}  \chi + \frac{\epsilon ^4}{36}  \left[  \Big(\pi {-}2 \text{Im}\left[\psi ^{(0)}\Big(\frac{i\chi}{\kappa}   {-}1\Big)\right]\Big)\chi  {-}(1{-}\kappa ) \left(2 \chi ^2{+}1\right)\right]} ~.
 \end{split}
\end{equation}
  Interestingly, even after using  identity \eqref{eq:identitypoly}, no-tree level contribution arises from the imaginary part of the near-zone. This signals dissipative effects arise purely as  loop-contributions. 

As a final remark, notice that from \eqref{eq:nears0l1m1}, no real term of the form $\chi^2\epsilon^3$ arises in the near-zone phase-shift. This is indeed corresponds to the vanishing of the  static leading Love number for $s=0$ perturbations off Kerr\footnote{Recall that Love numbers come from near-zone physics and have the scaling $\epsilon^{2\ell+1}$}.

\section{Appendix B: CFT Method Review }

In this appendix we briefly review the argument of  \cite{Bonelli:2022ten} to compute the CHE connection coefficients. Let us start by setting the notation: we consider Liouville CFT (for a review of Liouville theory, see \cite{Teschner:2001rv}), and parametrize the central charge as $c=1+6Q^2$, with $Q=b+b^{-1}$. We indicate primary operators of dimensions $\Delta_i = Q^2/4-\alpha_i^2$ as $V_{\alpha_i}(z_i)$, and the corresponding primary states as $|\Delta_i \rangle$. $\alpha_i$ is usually referred to as the Liouville momentum. Crucial for our discussion will be the so called rank 1 irregular state $\langle \mu, \Lambda |$ \cite{Gaiotto:2009ma,Bonelli:2011aa,
Gaiotto:2012sf}, which is defined as the state such that
\begin{equation}\label{eq:irrstate}
\langle \mu, \Lambda | L_0 = \Lambda \partial_\Lambda \,, \quad \langle \mu, \Lambda | L_{-1} = \mu \Lambda \langle \mu, \Lambda | L_{-1} \,, \quad \langle \mu, \Lambda | L_{-2} = -\frac{\Lambda^2}{4} \langle \mu, \Lambda | L_{-2} \,.
\end{equation}

\subsection{Connection Formula for CHE}
Let us consider the Liouville correlator 
\begin{equation}\label{eq:irrcorr}
    \langle \mu, \Lambda | V_{\alpha_1}(1) \Phi_{2,1}(z) |  \Delta_{\alpha_0} \rangle \,,
\end{equation}
where $\Phi_{2,1}$ is the level degenerate state of weight $\Delta_{2,1}=-\frac{1}{2}-\frac{3}{4}b^2$ that satisfies 
\begin{equation}\label{eq:BTZOP}
  \left(  b^{-2}L_{-1}^2+L_{-2}\right) \Phi_{2,1} (z) = 0 \,.
\end{equation}
Since Virasoro generators act as differential operators when inserted in correlation functions, equation \eqref{eq:BTZOP} turns into a differential equation for the correlator \eqref{eq:irrcorr}, that is the Belavin–Polyakov–Zamolodchikov (BPZ) equation \cite{Belavin:1984vu}
\begin{equation}
    \left(b^{-2}\partial_z^2 + \left(\frac{1}{z}+\frac{1}{z-1}\right) \partial_z + \frac{\Lambda \partial_\Lambda-\Delta_{2,1}-\Delta_0-\Delta_1}{z(z-1)} + \frac{\Delta_0}{z^2}+\frac{\Delta_1}{(z-1)^2} + \frac{\mu \Lambda}{z} - \frac{\Lambda^2}{4} \right) \langle \mu, \Lambda | V_{\alpha_1}(1) \Phi_{2,1}(z) |  \Delta_{\alpha_0} \rangle  = 0 \,.
\end{equation}
This is a partial differential equation in $\Lambda$ and $z$. In the semi-classical limit $ b \rightarrow 0, \alpha_i, \mu,\Lambda \rightarrow \infty $ such that $a_i = b \alpha_i, m_3 = b \mu, L = b \Lambda$ are finite, conformal blocks of \eqref{eq:irrcorr} behave as \cite{Zamolodchikov:1995aa}
\begin{equation}\label{eq:semiclcb}
\mathfrak{F} \sim \Lambda^{\Delta} \exp\left(\frac{F(a_1+a_0,a_1-a_0,m_3,a,L)}{b^2}+W(L;z)+ \mathcal{O}\left(b^{-2}\right)\right) \,,
\end{equation}
where $F$ is the so called classical conformal block and $\Delta= Q^2/4-\alpha^2$ ($a$ being the semiclassical momentum $\lim_{b\to0} b \alpha$) is the scaling dimension of the intermediate operator exchanged in the operator product expansion (OPE). The AGT correspondence \cite{Alday:2009aq} relates the classical Virasoro block $F(a_0+a_1,a_1-a_0,m_3,a,L)$\footnote{In the following we will suppress the dipendence of $F$ on $a_i,m_3$ to ease the notation.} to the instanton partition function of an $SU(2)$ $\mathcal{N}=2$ supersymmetric gauge theory with $N_f=3$ hypermultiplets of masses
\begin{equation}\label{eq:masses}
    m_1 = a_0 + a_1 ~, \quad m_2 = a_1 - a_0 ~, \quad m_3 = m_3 ~.
\end{equation}
in the NS phase of the $\Omega$ background. Besides its physical significance, the AGT correspondence gives a very convenient way of computing $F$ as we will see in the following.

Note that the $z-$dependence in \eqref{eq:semiclcb} enters at a subleading order in $b$, as one can expect from the fact that as $b$ goes to zero $\Delta_{2,1}$ is subleading with respect to $\Delta_i,\mu,\Lambda$. Crucially
\begin{equation}
    \Lambda \partial_\Lambda \mathfrak{F} = b^{-2}\left( \frac{1}{4}-a^2+ L \partial_L F(L) + \mathcal{O}\left(b^0\right) \right) \equiv b^{-2}\left( u + \mathcal{O}\left(b^0\right) \right) \,.
\end{equation}
The $\Lambda-$derivative decouples, leaving a new parameters, $u$, at its place. $u \equiv \frac{1}{4} - a^2$ is usually called the accessory parameter in the mathematical literature. All in all, semi-classical conformal  blocks defined as 
\begin{equation}
    \mathcal{F} = \lim_{b\to0} \Lambda^{-\Delta} e^{-\frac{1}{b^2}F} \mathfrak{F}
\end{equation}
satisfy the ODE
\begin{equation}
\label{eq:classical conformal block}
    \left(\partial_z^2+\frac{u-\frac{1}{2}+a_0^2+a_1^2}{z(z-1)}+\frac{\frac{1}{4}-a_1^2}{(z-1)^2}+\frac{\frac{1}{4}-a_0^2}{z^2}+\frac{m_3 L}{z}-\frac{L^2}{4}\right) \mathcal{F}(z) =0  ~,
\end{equation}
This ODE has two regular singularities at $z=0,1$ excited by the primary states, and an irregular singularity of rank 1 at $z=\infty$ generated by the irregular state: it is the CHE in its normal form.

The $z-$dependence of $\mathcal{F}(z)$ can be extracted by computing the OPE of the degenerate operator $\Phi_{2,1}(z)$ with the other insertions. When $\Phi_{2,1}(z)$ fuses e.g. with another primary one has
\begin{equation}\label{eq:OPEr}
    \Phi_{2,1}(z) V_{\alpha_i}(z_i) \sim \sum_\pm (z-z_i)^{\frac{1}{2}\pm a_i} \left(V_{\alpha_i \pm}(z_i)+\mathcal{O}(z-z_i) \right) \,.
\end{equation}
Inserting \eqref{eq:OPEr} into \eqref{eq:irrcorr} one can extract the $z$ dependence on the blocks for $z\sim z_i$. The $\pm$ signs corresponds for the two conformal dimensions exchanged by the OPE, and accounts for the two linearly independent local solutions of the ODE. More precisely, one  finds 
around $z=1$
\begin{equation}
    \mathcal{F}_{\pm}^{(1)}(1-z) = e^{\mp \frac{1}{2} \partial_{a_1} F} (1-z)^{\frac{1}{2}+\pm a_1} ~,
\end{equation}
and around $z=\infty$ 
\begin{equation}
    \mathcal{F}_{\pm}^{(\infty)}(1/z) = e^{\mp \frac{1}{2} \partial_{m_3} F} e^{\pm L z/2} L^{-\frac{1}{2} \mp m_3} z^{\mp m_3} ~.
\end{equation}

The connection formula can be worked out by using crossing symmetries in the conformal correlators. We refer the reader to \cite{Bonelli:2022ten} for more details, and just sketch the main idea here. Crossing symmetry relates between each other different OPE decomposition of the correlator \eqref{eq:irrcorr}. As mentioned above different OPE decompositions reconstruct local solution of the ODE centered close to different singular points. Schematically, crossing symmetry constraint take the following form:
\begin{equation}\label{eq:crossing}
\sum_{\pm} (\text{3 pt functions}) \big| \mathfrak{F}^{(1)}_\pm (1-z) \big|^2 = \sum_{\pm} (\text{3 pt functions}) \big| \mathfrak{F}^{(\infty)}_\pm (z^{-1}) \big|^2 \,.
\end{equation}
The 3 point functions of Liouville CFT are non-perturbatively known \cite{Dorn:1994xn,Zamolodchikov:1995aa}. One can then use \eqref{eq:crossing} to express e.g. $\mathfrak{F}^{(1)}_+ (1-z)$ as a linear combination of $\mathfrak{F}^{(\infty)}_\pm (z^{-1})$. Upon taking the semi-classical limit, this allows use to compute the connection coefficients of the CHE. For the relevance of this paper, we quote the formula
\begin{equation}
    \mathcal{F}_{\theta}^{(1)}(1-z) = \sum_{\theta' = \pm} \mathcal{M}(\theta, \theta') \mathcal{F}_{\theta'}^{(\infty)}(1/z) ~, \quad \theta = \pm ~.
\end{equation}
with the connection matrix
\begin{equation}
\label{eq:semi-block connection}
    \mathcal{M}(\theta,\theta') =\sum_{\sigma= \pm}  L^{\sigma a} \frac{\Gamma(1-2 \sigma a) \Gamma(-2 \sigma a) \Gamma\left(1+2 \theta a_1\right) }{\Gamma\left(\frac{1}{2}+\theta a_1-\sigma a+a_0\right) \Gamma\left(\frac{1}{2}+\theta a_1-\sigma a-a_0\right) \Gamma\left(\frac{1}{2}-\sigma a-\theta^{\prime} m_3 \right)} e^{i \pi\left(\frac{1-\theta^{\prime}}{2}\right)\left(\frac{1}{2}-m_3-\sigma a\right)} e^{-\frac{\sigma}{2} \partial_a F} ~.
\end{equation}

\subsection{Solving Radial TME}
Now, we apply the CFT method to solving the radial TME satisfied by ${}_s R_{\ell m}(r)$ in \eqref{eq:TME}. Performing the following changing of variables
\begin{equation}
\Psi(z) = \Delta^{\frac{s+1}{2}} (r) {}_s R_{\ell m \omega}(r) \,, \quad z = \frac{r-r_-}{r_+-r_-} \,,
\end{equation}
the TME takes the form \eqref{eq:classical conformal block}, the in-going solution to radial TME at the horizon can be written as
\begin{equation}
\label{eq:CFT in solution}
\Psi^{\rm in}(z) = {}_sC^-_{\ell m} e^{\frac{L z}{2}} z^{- m_3} \left(1+\mathcal{O}(z^{-1)}\right) + {}_sC^+_{\ell m} e^{-\frac{L z}{2}} z^{m_3} \left(1+\mathcal{O}(z^{-1)}\right)
\end{equation}
where ${}_s C^\pm_{\ell m}$ are elements of the Heun connection matrix \eqref{eq:semi-block connection}. Explicitly,
\begin{equation}
\begin{aligned}
&{}_sC^-_{ \ell m } = \sum_{ \sigma = \pm} L^{-\frac{1}{2} - m_3 + \sigma a} \frac{ \Gamma\left(1-2\sigma a \right) \Gamma\left(-2\sigma a \right)\Gamma\left(1+2a_1\right) }{ \Gamma\left(\frac{1}{2}+a_1-\sigma a+a_0\right)\Gamma\left(\frac{1}{2}+a_1-\sigma a-a_0\right)\Gamma\left(\frac{1}{2}-\sigma a-m_3\right)} e^{-\frac{\sigma}{2} \partial_a F-\frac{1}{2} \partial_{m_3} F} \,, \\
&{}_sC^+_{ \ell m} = \sum_{ \sigma = \pm} (-L)^{-\frac{1}{2} + m_3 + \sigma a} \frac{ \Gamma\left(1-2\sigma a\right) \Gamma\left(-2\sigma a\right)\Gamma\left(1+2a_1\right) }{ \Gamma\left(\frac{1}{2}+a_1-\sigma a +a_0\right)\Gamma\left(\frac{1}{2}+a_1-\sigma a-a_0\right)\Gamma\left(\frac{1}{2}-\sigma a+m_3\right)} e^{-\frac{\sigma}{2} \partial_a F+\frac{1}{2} \partial_{m_3} F} \,.
\end{aligned}
\end{equation}
Mapping \eqref{eq:CFT in solution} to the asymptotic behavior given in \eqref{eq:bdary}, we get \eqref{eq:nearpfarCFT} in the main text.

\subsection{Computation of $F$ and Symmetry Properties}
As shown in \eqref{eq:semiclcb}, $F$ controls the $z-$independent part of the correlator \eqref{eq:irrcorr}, that is
\begin{equation}\label{eq:Fblock}
    \langle \mu,\Lambda | V_{\alpha_1}(1) | \Delta_0 \rangle \,.
\end{equation}
Small-$\Lambda$ conformal blocks of \eqref{eq:Fblock} are given by 
\begin{equation}\label{eq:frakF}
    \mathfrak{F} \left(\mu, \alpha_i, \alpha, \Lambda \right) = \Lambda^\Delta e^{\left( \frac{Q}{2} + \alpha_1 \right) \Lambda}  \sum_{\vec{Y}} \Lambda^{| \vec{Y} |} z_{\text{vec}} \left( \vec{\alpha}, \vec{Y} \right) z_{\text{hyp}} \left( \vec{\alpha}, \vec{Y}, - \mu \right) \prod_{\theta = \pm} z_{\text{hyp}} \left( \vec{\alpha}, \vec{Y}, \alpha_1 + \theta \alpha_0 \right) \,.
\end{equation}
where the sum runs over pairs of Young tableaux $\left( Y_1, Y_2 \right)$, and $\alpha$ is the Liouville momentum of the intermediate operator exchanged in the $V_1(1) | \Delta_0 \rangle$ OPE. We denote the size of the pair $| \vec{Y} | = | Y_1 | + | Y_2 |$, and
\cite{Flume:2002az,Bruzzo:2002xf}
\begin{equation}\label{zdef}
\begin{aligned}
    &z_{\text{hyp}} \left( \vec{\alpha}, \vec{Y}, \mu \right) = \prod_{k= 1,2} \prod_{(i,j) \in Y_k} \left( \alpha_k + \mu + b^{-1} \left( i - \frac{1}{2} \right) + b \left( j - \frac{1}{2} \right) \right) \,, \\
    &z_{\text{vec}} \left( \vec{\alpha}, \vec{Y} \right) = \prod_{k,l = 1,2} \prod_{(i,j) \in Y_k}  E^{-1} \left( \alpha_k - \alpha_l, Y_k, Y_l, (i, j) \right) \prod_{(i',j') \in Y_l} \left( Q - E \left( \alpha_l - \alpha_k, Y_l, Y_k, (i', j') \right) \right)^{-1}\,, \\
    &E \left( \alpha, Y_1, Y_2, (i,j) \right) = \alpha - b^{-1} L_{Y_2} ((i,j)) + b \left( A_{Y_1} ((i,j)) + 1 \right) \,.
\end{aligned}
\end{equation}
Here $L_Y ((i,j)), A_Y((i,j))$ denote respectively the leg-length and the arm-length of the box at the site $(i,j)$ of the tableau $Y$. If we denote a Young tableau as $Y = ( \nu_1' \ge \nu_2' \ge \dots)$ and its transpose as $Y^T = ( \nu_1 \ge \nu_2 \ge \dots)$, then $L_Y$ and $A_Y$ read
\begin{equation}
    A_Y (i, j) = \nu_i' -  j \,, \, \, L_Y (i, j) = \nu_j - i \,.
\end{equation}
\begin{figure}[t!]\centering
\includegraphics[scale=0.5]{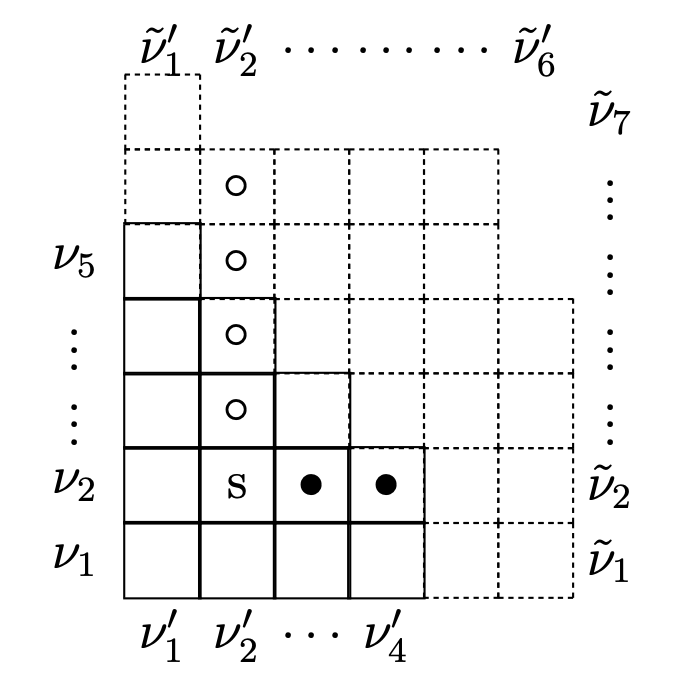}
\caption{Arm length $A_{\tilde{Y}} (s)=4$ (white circles) and leg length $L_Y(s)=2$ (black dots) of a box at the site $s = (2,2)$ for the pair of superimposed diagrams $Y$ (solid lines) and $\tilde{Y}$ (dotted lines).}
\end{figure}
Note that they can be negative if the box $(i,j)$ are the coordinates of a box outside the tableau. Also, the previous formulae has to be evaluated at $\vec{\alpha} = (\alpha_1, \alpha_2) = (\alpha, - \alpha)$.
The explicit expression for the NS function, a.k.a. classical Virasoro confluent conformal block $F$, is finally given by :
\begin{equation}\label{eq:semiclassFfromF}
     F(L) = \lim_{b \to 0} b^2 \log
     \Lambda^{-\Delta} \mathfrak{F} \left(m_3/b, a_i/b, a/b, L/b \right) 
\end{equation}
Finally, let us comment on the symmetry properties discussed around Eq. \eqref{eq:f-tilde}. We start by proving that $F$ is invariant under $(m_3,L)\to(-m_3,-L)$. $F$ is defined in terms of \eqref{eq:Fblock}, and the dependence on $(\mu,\Lambda)$ is entirely controlled by the irregular state. From \eqref{eq:irrstate} we see that the irregular state is invariant under $(\mu,\Lambda)\to(-\mu,-\Lambda)$, so the same must be true for the whole correlator. This property descends to the conformal blocks \eqref{eq:Fblock}. In the semiclassical limit \eqref{eq:semiclassFfromF} this proves invariance of $F$ under $(m_3,L)\to(-m_3,-L)$. 

We now prove symmetry properties of $F$ under permutation of masses. First of all note that 
\begin{equation}
    \sum_{\vec{Y}} \Lambda^{| \vec{Y} |} z_{\text{vec}} \left( \vec{\alpha}, \vec{Y} \right) z_{\text{hyp}} \left( \vec{\alpha}, \vec{Y}, - \mu \right) \prod_{\theta = \pm} z_{\text{hyp}} \left( \vec{\alpha}, \vec{Y}, \alpha_1 + \theta \alpha_0 \right)
\end{equation}
is symmetric under permutations  of $(\alpha_1+\alpha_0,\alpha_1-\alpha_0,\mu)$. The only non symmetric term in \eqref{eq:frakF} is the overall exponential term. However
\begin{equation}
\tilde{\mathfrak{F}} = e^{-\frac{1}{2}\mu\Lambda} \mathfrak{F}
\end{equation}
has the permutation symmetry. Upon taking the semiclassical limit \eqref{eq:semiclassFfromF}, this proves that $\tilde{F}$ as defined in \eqref{eq:f-tilde} is symmetric under permutations of masses. Combining this property with symmetry under $(m_3,L)\to(-m_3,-L)$, finally proves that $\tilde{F}$ is symmetric under $(m_i,L)\to(-m_i,-L)$ for $i=1,2,3$.

\section{Appendix C: Proof of Large Frequency Behavior}

We now present a CFT argument to prove the fact that at large $\omega$, the renormalized angular momentum becomes
$a \approx 2iM\omega$ as indicated in \eqref{eq: analytic estimation nu} in the main text. We start from the correlator
\begin{equation}\label{eq:invariantcorre1}
e^{-\frac{1}{2} \Lambda\mu} \langle \mu, \Lambda | V_{\alpha_1}(1) | \Delta_{\alpha_0} \rangle \,
\end{equation}
Exchanging $\mu$ and $\alpha_0+\alpha_1$ gives
\begin{equation}\label{eq:invariantcorre2}
e^{-\frac{1}{2} \Lambda (\alpha_1+\alpha_0)}\langle (\alpha_1+\alpha_0), \Lambda | V_{\tilde{\alpha}_1}(1) | \Delta_{\tilde{\alpha}_0} \rangle \,.
\end{equation}
where 
\begin{equation}
\tilde{\alpha_1} = \frac{\mu_3+\alpha_1-\alpha_0}{2} \,, \quad \tilde{\alpha_0} = \frac{\mu_3-\alpha_1+\alpha_0}{2} \,.
\end{equation}
Note that \eqref{eq:invariantcorre1} and \eqref{eq:invariantcorre2} define the same $\tilde{F}$. Defining as usual $\tilde{a}_{1,2} = \lim_{b\to 0} b \tilde{\alpha}_{1,2}$, the dictionary at large $\omega$ gives at leading order.
\begin{equation}
    a_1+a_0 \simeq \frac{-2iM\omega}{\kappa} \,, \quad \tilde{a}_0 \simeq 2 i M \omega \,, \quad \tilde{a}_1 \simeq \mathcal{O}(1) \,.
\end{equation}
$\tilde{a}_1$ is subleading with respect to the other parameters, therefore one can neglect the insertion at $1$ in \eqref{eq:invariantcorre2}. This gives 
\begin{equation}
    e^{-\frac{1}{2} \Lambda (\alpha_1+\alpha_0)}\langle (\alpha_1+\alpha_0), \Lambda | V_{\tilde{\alpha}_1}(1) | \Delta_{\tilde{\alpha}_0} \rangle \simeq e^{-\frac{1}{2} \Lambda (\alpha_1+\alpha_0)}\langle (\alpha_1+\alpha_0), \Lambda | \Delta_{\tilde{\alpha}_0} \rangle = e^{-\frac{1}{2} \Lambda (\alpha_1+\alpha_0)} \Lambda^\Delta \,,
\end{equation}
where we used the fact that $\langle \mu, \Lambda | \Delta_i \rangle = \Lambda^{\Delta_i}$. This gives for $\tilde{F}$
\begin{equation}\label{eq:ftildapp}
\begin{aligned}
&\tilde{F} \simeq = -\frac{L}{2} (a_1+a_0)  \,,
\end{aligned}
\end{equation}
Inserting this in the Matone relation \eqref{eq:Matone} we find the solution $a \simeq m_3$ (fixing to $+$ the $\pm$ ambiguity coming from the fact that \eqref{eq:Matone} is quadratic in $a$). As a consistency condition note that this gives for $u$
\begin{equation}
u \simeq \frac{1}{2}-m_3^2 + \frac{L}{2}(m_3-m_1) \,,
\end{equation}
consistently with the large $\omega$ limit of \eqref{eq:dict}.
Using \eqref{eq:masses} into \eqref{eq:ftildapp} we finally recover \eqref{eq:Fhfl} in the main text.

\end{document}